\documentclass{jfm}
\usepackage{amsmath}
\usepackage{graphicx}
\usepackage{dcolumn}
\usepackage{bm}
\usepackage{hyperref}
\usepackage{lineno}
\usepackage{color}
\usepackage{tikz}
\usepackage{pgfplots}
\usetikzlibrary{shapes}
\usepackage{amssymb}
\usepackage{appendix}

\def\XXint#1#2#3{{\setbox0=\hbox{$#1{#2#3}{\int}$}
     \vcenter{\hbox{$#2#3$}}\kern-.5\wd0}}

\usepackage{fancyhdr} 
\shorttitle{Reduced equations for strongly stratified flows}
\shortauthor{G. P. Chini and G. Michel}

\title{Exploiting self-organized criticality in strongly stratified turbulence}

\author{Gregory P. Chini\aff{1,2}\corresp{\email{greg.chini@unh.edu}},
  Guillaume Michel\aff{3},
  Keith Julien\aff{4},
  Cesar B. Rocha\aff{5},
  \and  Colm-cille P. Caulfield\aff{6}
}
\affiliation{\aff{1} Program in Integrated Applied Mathematics, University of New Hampshire, Durham, NH 03824, USA
\aff{2} Department of Mechanical Engineering, University of New Hampshire, Durham, NH 03824, USA
\aff{3} Institut Jean Le Rond d'Alembert, Sorbonne Universit\'e, CNRS, UMR 7190 Paris, F-75005, France
\aff{4} Department of Applied Mathematics, University of Colorado, Boulder, CO 80309, USA
\aff{5} Department of Marine Sciences, University of Connecticut, Storrs, CT 06269, USA
\aff{6} BP Institute and Department of Applied Mathematics \& Theoretical Physics, Cambridge University, Cambridge, CB3 0WA UK}

\begin{document}
\maketitle
\begin{abstract}
A multiscale reduced description of turbulent free shear flows in the presence of strong stabilizing density stratification is derived via asymptotic analysis of the Boussinesq equations in the simultaneous limits of small Froude and large Reynolds numbers.  The analysis explicitly recognizes the occurrence of dynamics on disparate spatiotemporal scales, yielding simplified partial differential equations governing the coupled evolution of slow large-scale hydrostatic flows and fast small-scale isotropic instabilities and internal waves.  The dynamics captured by the coupled reduced equations is illustrated in the context of two-dimensional strongly stratified Kolmogorov flow.  
A noteworthy feature of the reduced model is that the fluctuations are constrained to satisfy quasilinear (QL) dynamics about the comparably slowly-varying large-scale fields.  Crucially, this QL reduction is not invoked as an \textit{ad hoc} closure approximation, but rather is derived in a physically relevant and mathematically consistent distinguished limit. Further analysis of the resulting slow-fast QL system shows how the amplitude of the fast stratified-shear instabilities is slaved to the slowly-evolving mean fields to ensure the marginal stability of the latter. Physically, this marginal stability condition appears to be compatible with recent evidence of self-organized criticality in both observations and simulations of stratified turbulence. Algorithmically, the slaving of the fluctuation fields enables numerical simulations to be time-evolved strictly on the slow time scale of the hydrostatic flow. The reduced equations thus provide a solid mathematical foundation for future studies of three-dimensional strongly stratified turbulence in extreme parameter regimes of geophysical relevance and suggest avenues for new sub-grid-scale parameterizations.
\end{abstract}

\section{Introduction}\label{sec:INTRO}
\begin{figure}
\begin{center}
\includegraphics[width=0.99\linewidth]{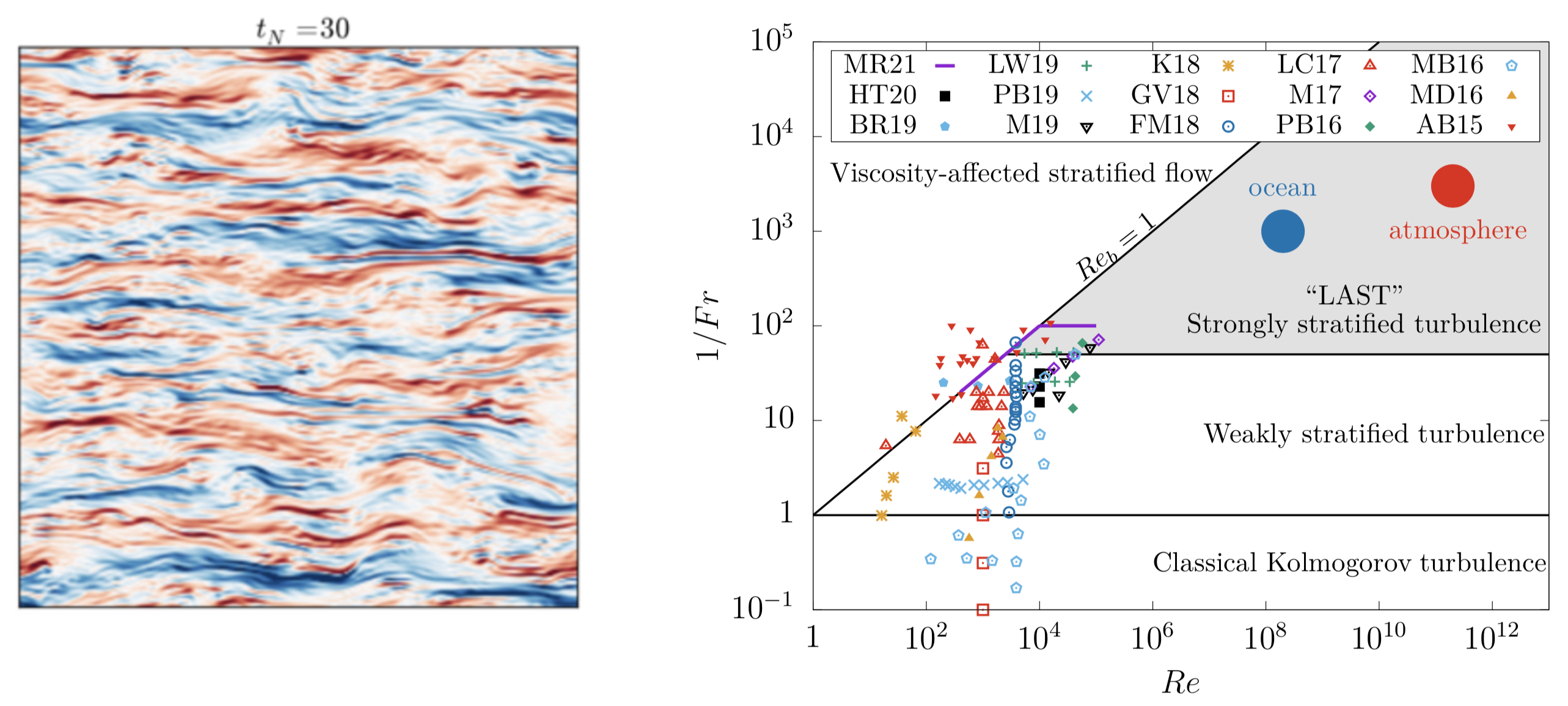}
\end{center}
\caption{Physical attributes and parameter regime characterizing stratified turbulence. (a)~Snapshot of the vorticity field in a freely-decaying two-dimensional (2D) DNS of the Boussinesq equations (in a doubly-periodic horizontal/vertical domain) for $Pr=1$, $Fr=0.02$ and $Re=5\times 10^5$ after 30 buoyancy time units $2\pi/N$ (i.e. $t_N=30$), where $N$ is the constant buoyancy frequency, initialized with a random vorticity field having a von-Karman-like spectrum.
Note the emergence of highly anisotropic layers, with horizontal scales $L$ much greater than their vertical thickness $h$, coexisting with small-scale, roughly isotropic variability suggestive of Kelvin--Helmholtz overturns. (b)~Regime diagram adapted from \cite{brethouwer2007}. The 2D DNS and reduced multiscale simulations performed in the present work lie along the purple solid line, and typical values for the upper ocean and middle atmosphere are taken from \cite{Moum_1996} and \cite{Lindborg_2006}. DNS published during the past six years are also reported: `HT20' \citep{Howland_2020}, `BR19' \citep{deBruynKops_Riley_2019}, `LW19' \citep{Lang_2019}, `PB19' \citep{Portwood_2019}, `M19' \citep{Maffioli_2019}, `K18' \citep{Khani_2018}, `GV18' \citep{Garanaik_2018}, `FM18' \citep{Feraco_2018}, `LC17' \citep{Lucas_2017_2}, `M17' \citep{Maffioli_2017}, `PB16' \citep{Portwood_2016}, `MB16' \citep{Maffioli_2016_2}, `MD16' \citep{Maffioli_2016}, `AB15' \citep{Augier_2015}. {{Note that there}} is slight variability in the precise definitions of both $Fr$ and $Re$ for certain data collected in this regime diagram {{and that there are not sharp transitions among the various regimes.}}}
\label{RegimeDiagram}
\end{figure}

Strongly stratified turbulent flows occur routinely both in the natural and built environment.  In many geoscientific and technological applications, these flows exert a controlling influence on the turbulent mixing of buoyancy, momentum and mass, yet numerous fundamental questions concerning the structure and mechanics of stratified mixing in extreme parameter regimes remain open.  Owing to the spontaneous emergence of both small-aspect-ratio hydrostatic flow structures having large horizontal scales and roughly isotropic, non-hydrostatic small-scale instabilities and waves (see figure~\ref{RegimeDiagram}\textit{a}), an enormous range of spatiotemporal scales must be resolved. This scale disparity renders measurements and direct numerical simulations (DNS) of such strongly stratified turbulence, referred to as the \emph{layered anisotropic stratified turbulence} (LAST) regime by \cite{FalderWhiteCaulfield_JPO_2016}, especially challenging.  In the oceans and atmosphere, for example, non-rotating stratified turbulence is considered to be the prevailing dynamics on horizontal length scales $L$ smaller than that of the large-scale rotationally-constrained (quasi-geostrophic) flow and greater than the Ozmidov scale $L_O\equiv\sqrt{\epsilon_h/N^3}$, below which buoyancy forces are negligible. Here, $\epsilon_h$ and $N$ are the turbulent kinetic energy dissipation rate and the buoyancy frequency, respectively.  Presuming that $\epsilon_h\sim U^3/L$, where $U$ is a horizontal velocity characterizing motions at horizontal scale $L$, it is straightforward to show that the ratio $L/L_O=\mathit{O}(Fr^{-3/2})$, where the (horizontal) Froude number $Fr\equiv U/(NL)$.  Thus, in the limit of strong stratification, a large range of scales must be resolved.  Indeed, when the Ozmidov scale is much larger than the Kolmogorov scale, two inertial ranges in principle must be captured.  Collectively, these attributes render DNS of strongly stratified turbulence arduous even with state-of-the-art supercomputing power.

Away from boundaries and for fixed $\mathit{O}(1)$ Prandtl number $Pr$, stratified turbulence is governed by two external control parameters:  the Froude number $Fr$ and the Reynolds number $Re\equiv U L/\nu$, where $\nu$ is the kinematic viscosity.  The challenges associated with accessing the LAST regime in the laboratory or via DNS of the governing Boussinesq equations are aptly described by \cite{brethouwer2007}.  In particular, figure~\ref{RegimeDiagram}\textit{b}, an adaptation of their regime diagram, confirms that recent DNS (i.e. simulations documented in the past six years employing algorithms devoid of hyperviscosity or any other \textit{ad hoc} small-scale modeling) have been performed in a parameter regime orders of magnitude away from that characterizing the LAST regime of strongly stratified turbulence as it is believed to occur in the oceans and atmosphere, where $Re>10^8$ and $Fr<10^{-3}$ (based on `outer' turbulence scales).  \cite{Bartello_2013} estimate that to employ DNS to settle a variety of fundamental scientific questions regarding the dynamics of strongly stratified turbulence would require the ratio of the maximum to minimum resolvable scale to be in the millions (i.e. in a single coordinate direction), yielding a formidable computational challenge.  These questions include, for example, 
what is the mixing efficiency of strongly stratified turbulence in the LAST regime in the small-$Fr$, large-$Re$ limit?  Is this efficiency dependent upon the precise way in which these limits are taken?  Does the resulting efficiency depend on the details of the initial stratification?  Is the horizontal spectrum of horizontal kinetic energy independent of $Fr$ as $Fr\to 0$ with $Re>10/Fr^2$, as conjectured by some previous investigators~\citep{Lindborg_2006}? What is the physical origin of the tendency for strongly stratified turbulence to exhibit attributes of self-organized criticality \citep{SmythGRL2013,GeyerJPO2016,SalehipourJFM2018,SmythNature2019}? To date, these and other important questions have not been conclusively answered.

In contrast to DNS, regional oceanic and atmospheric numerical circulation models do not attempt to resolve the small-scale dynamics of stratified mixing.  Indeed, the fluid dynamical core of these computational models generally comprises the so-called \textit{hydrostatic} primitive equations \citep{Miller_2007,Klein_2010}.  
Nevertheless, small-scale non-hydrostatic stratified mixing events---often although not exclusively associated with the breakdown of internal waves \citep{Legg_ARFM_2021,MacKinnon_BAMS_2017}---ultimately exert strong feedbacks on the resolved large-scale flow, particularly to close the global buoyancy budget \citep{Ferrari_Wunsch_ARFM_2004,Gregg_2018}; therefore, suitable parameterizations must be developed and implemented.
Generally, these parameterizations are physically motivated but ultimately \textit{ad hoc} and thus raise the attendant issues over robustness, accuracy and reliability.  For ocean simulations, for example, \cite{Gregg_2018} recently concluded that a constant mixing efficiency equal to 0.2 should continue to be used as a parametrization until a better understanding of stratified turbulence is achieved, despite significant discrepancies with \textit{in situ} measurements of stratified mixing. (See \cite{Monismith_2018} for an additional description of strongly stratified turbulence field data.) Even the sub-grid scale (SGS) models employed in large-eddy simulations (LES) are notoriously suspect in strongly stratified turbulence owing to the high degree of anisotropy of the dominant flow structures and the complex patterns of backscatter of energy from small to large scales. The recent numerical study by \cite{Khani_2018} has shown that, even for dimensionless parameters orders of magnitude different from those realised in nature, the mixing efficiency obtained from LES agrees with that computed from fully resolved DNS only when the spatial discretization used in the LES is comparable to or smaller than $L_O$. Consequently, LES must resolve a scale separation that is at least $\mathit{O}(Fr^{-3/2})$ to be reliable.

One intriguing aspect of strongly stratified turbulence is its tendency to exhibit aspects of \textit{self-organized criticality}, particularly the condition of marginal stability. Indeed, mounting evidence suggests an apparent connection between the gradient Richardson number observed in statistically-steady (or at least slowly-varying) stratified turbulence and the celebrated Miles--Howard criterion for linear instability of inviscid \textit{laminar} stratified shear flows.  For example, \cite{SmythGRL2013} analyzed five years of data taken from the upper 150 meters of the eastern equatorial Pacific Ocean; their analysis clearly shows that in a depth range from 20 to 80 meters, in which diurnally-cycling stratified turbulence is evident, the probability density function (pdf) of the measured gradient Richardson number peaks at a value very close to 1/4. The authors argue that these measurements indicate deep-cycle stratified turbulence in the equatorial Pacific Ocean occurs about a mean state that is close to marginally stable and that the phenomenon is an example of self-organized critical behavior.  Further discussion is provided by \cite{SmythNature2019}.  Similarly, the field measurements of \cite{GeyerJPO2016} in a salt-stratified estuary (the Connecticut River) show that, away from the river bottom, the tidally-driven, quasi-steady stratified turbulence is characterized by a median gradient Richardson number equal to 0.25. Perhaps the most compelling evidence to date of the potential for strongly-stratified turbulence to exhibit self-organized criticality is provided by \cite{SalehipourJFM2018}, who carried out DNS of stratified free shear layers in which there is a region of substantially enhanced density gradient embedded within the shear zone.  Under these conditions, stratified turbulence is engendered by the breakdown of Holmboe instability waves. The authors show that this instability drives long-lived, quasi-stationary turbulent mixing events for which the pdf of the instantaneous, horizontally-averaged gradient Richardson number again has a maximum very close to 0.25. Although not of central relevance to the present investigation, \cite{SalehipourJFM2018} also demonstrate that the resulting stratified turbulence exhibits other aspects of self-organized criticality, including scale-invariant `avalanches' of small-scale turbulence. The authors argue that the emergence of a mean turbulence state with a gradient Richardson number approximately equal to one-quarter is not coincidental but instead is inherently connected to the intrinsic self-regulation of the strongly-stratified flow dynamics.

In this study, we develop a systematically simplified mathematical framework for theoretical and computational investigations of strongly stratified free shear flows that enables simulations in extreme parameter regimes of physical relevance while largely ameliorating the need for phenomenological modeling.  Our approach is inspired by the work of \cite{kJ07}, who demonstrate that multiple-scales asymptotic analysis can be used to derive reduced equations for turbulent flows subjected to external constraints such as rapid system rotation or a strong magnetic field.  Heuristically, the strong constraint inhibits mode coupling in particular spatial directions and is associated with the emergence of highly anisotropic flow structures (e.g. turbulent Taylor columns in the rapid rotation scenario).  Mathematically, particular dominant balances of terms in the governing equations (e.g. geostrophic balance) arise in extreme parameter regimes, and reduced equations can be derived by systematically perturbing about this balanced state.  

In our work, the turbulence is strongly constrained by the imposed stabilizing stratification, but the Reynolds number is sufficiently large that the laminar state can be destabilized.  We exploit this constraint to derive in the physically-relevant asymptotic limits of small Froude and large Reynolds numbers a closed reduced set of partial differential equations that captures the leading-order coupled dynamics of the large-scale anisotropic hydrostatic flow and small-scale non-hydrostatic instabilities and waves characterized by their commensurate horizontal and vertical length scales.  The dynamics admitted by the coupled reduced equations is illustrated in the context of two-dimensional (2D) strongly stratified Kolmogorov flow.  
A noteworthy feature of the reduced model is that the fluctuations are constrained to satisfy quasilinear (QL) dynamics about the comparably slowly-varying large-scale fields.  Crucially, this QL reduction is not invoked as an \textit{ad hoc} closure approximation, e.g. as in \cite{Fitzgerald_2014,Fitzgerald_2018}, but rather is derived here in a physically relevant and mathematically consistent distinguished limit.  The reduced equations thus provide a solid mathematical foundation for future studies of 
three-dimensional strongly stratified turbulence (i.e. of LAST) in extreme parameter regimes of geophysical relevance. 

In subsequent sections, we first derive reduced equations for large Reynolds number strongly-stratified flow using multiple scales analysis (\S~\ref{ASYMPTOTICS}). We then introduce a novel scheme for integrating the resulting slow--fast QL system strictly on the slow time scale characterizing the mean-field evolution (\S~\ref{ALGORITHM}). In \S~\ref{RESULTS}, we apply this new framework to the computation of exact coherent states (ECS) in strongly stratified Kolmogorov flow as an illustrative proof of concept of the utility of the reduced equation set. Finally, in \S~\ref{CONCLUSION}, we draw our conclusions and propose future avenues of research.

\section{Multiple scales reduction of the Boussinesq equations}\label{ASYMPTOTICS}

We consider a volume of stably stratified fluid far from any boundaries that is driven by an imposed body force. Density (or buoyancy) variations are incorporated through the Boussinesq approximation.  In the LAST regime, it is well established that anisotropic flow structures emerge with horizontal scales $L\gg h$ \citep{Riley_2000,Riley_2013}, where $h$ is the typical vertical scale of variation of the dependent field variables (see figure~\ref{RegimeDiagram}\textit{a}).  Thus, we non-dimensionalize the governing Boussinesq equations anisotropically, scaling horizontal velocities with $U$, horizontal distance with $L$, time with $L/U$ and pressure with $\rho_0 U^2$, where $\rho_0$ is a constant reference density, whereas vertical distance is scaled with $h$, vertical velocities with $Fr^2 (L/h) U$ and buoyancy (more precisely, the negative reduced gravity) with $U^2/h$.  The buoyancy scale is deduced by ensuring that hydrostatic balance can be attained on large horizontal scales, while, in the first instance, the scaling for the vertical velocity arises from balancing horizontal and vertical advection of buoyancy rather than from imposing three-dimensional (3D) incompressibility on those scales.   The resulting dimensionless Boussinesq system can be expressed as
\begin{eqnarray}
\partial_t\mathbf{u}_\bot+\left(\mathbf{u}_\bot\cdot\nabla_\bot\right)\mathbf{u}_\bot+{\frac{Fr^2}{\alpha^2}}\,W\partial_z\mathbf{u}_\bot &=& -\nabla_\bot p+{\frac{1}{Re}}\left[\nabla_\bot^2+{\frac{1}{\alpha^2}}\partial_z^2\right]\mathbf{u}_\bot+ \mathbf{f}_\bot,\label{eqn:Bu}\\
{Fr}\left[\partial_t W + \left(\mathbf{u}_\bot\cdot\nabla_\bot\right)W+{\frac{Fr^2}{\alpha^2}}W\partial_z W\right] &=& \frac{1}{Fr}\left(-\partial_z p + b\right) +{\frac{Fr}{Re}}\left[\nabla_\bot^2+{\frac{1}{\alpha^2}}\partial_z^2\right]W,\label{eqn:Bw}\\
\nabla_\bot\cdot\mathbf{u}_\bot + {\frac{Fr^2}{\alpha^2}}\partial_z W &=&  0,\label{eqn:Bcont}\\
\partial_t b + \left(\mathbf{u}_\bot\cdot\nabla_\bot\right)b + {\frac{Fr^2}{\alpha^2}}W\partial_z b &=& -W + {\frac{1}{Pr Re}}\left[\nabla_\bot^2+{\frac{1}{\alpha^2}}\partial_z^2\right]b.\label{eqn:Bb}
\end{eqnarray}
{{In (\ref{eqn:Bu})--(\ref{eqn:Bb}), $\alpha\equiv h/L$ is the aspect ratio characterizing the anisotropy of typical large scale flow structures, and the subscript $\bot$ denotes the horizontal ($x$,$y$) plane, i.e. the plane perpendicular to gravity.}}
The velocity vector $\mathbf{u}$=($\mathbf{u}_\bot$,$W$), where $\mathbf{u}_\bot$=($u$,$v$) is the horizontal velocity vector and $W$ is the vertical ($z$) velocity component; $b$ is the buoyancy deviation from an {{imposed locally linearly-varying background profile (i.e. which varies on a much larger, dimensional scale height comparable to $L$)}} so that the total buoyancy field is given by $z+b$; and $p$ is the pressure. A body force $\mathbf{f}_\bot$ is incorporated into the horizontal momentum equations to drive the flow.

To obtain limiting equations governing the dynamics in the strongly stratified regime, a determination must be made regarding the behavior of the flow aspect-ratio $\alpha$ in the limit $Fr\to 0$.  \cite{Lilly_1983} assumed that $\alpha$ remains fixed in that limit, in which case the reduced version of (\ref{eqn:Bu})--(\ref{eqn:Bb}) governs layerwise -- that is, essentially vertically decoupled -- 2D dynamics.  If, however, $\alpha=\mathit{O}(Fr)$ as $Fr\to 0$, as proposed later by \cite{Billant_2001}, then (\ref{eqn:Bu})--(\ref{eqn:Bb}) reduce to the so-called hydrostatic primitive equations \citep{Riley_2013} governing strongly anisotropic but nevertheless 3D turbulence. [Interestingly, to the best of our knowledge, the compatibility of this scaling with the requirements for the self-consistency of the Boussinesq approximation, e.g. see \cite{Bois_1991}, has never been analysed.]  A spate of subsequent investigations has empirically confirmed the relevance of the Billant--Chomaz scaling \citep{Waite_2004,Lindborg_2006,brethouwer2007,Augier_2012,Augier_2015,Maffioli_2016,Lucas_2017}.  Physically, the implication is that the as yet unspecified vertical length scale $h$ of the large horizontal-scale flow structures dynamically self-adjusts to be $\mathit{O}(U/N)$, the so-called buoyancy scale; i.e. the \textit{vertical} Froude number $U/(Nh)=\mathit{O}(1)$ as $Fr\to 0$.  In the following analysis, we presume the Billant--Chomaz scaling is the relevant one for strongly stratified turbulence; i.e. we assume that the turbulence evolves to be in this inherently vertically layered and anisotropic state.

Although the hydrostatic primitive equations capture vertical mode coupling in the LAST regime, they necessarily fail to include small-scale, isotropic and non-hydrostatic dynamics, the effects of which usually must be phenomenologically modeled, as noted above.  Here, we proceed more systematically by formally introducing `fast' horizontal and temporal independent variables, $\boldsymbol{\chi}_\bot\equiv \mathbf{x}_\bot/Fr$ and $\tau\equiv t/Fr$, so that derivatives transform as
\begin{eqnarray}
\nabla_\bot\to (1/Fr)\nabla_{\boldsymbol{\chi}_\bot}+\nabla_{\mathbf{x}_\bot}\;\mbox{and}\;\;\partial_t\to (1/Fr)\partial_\tau+\partial_t
\end{eqnarray}
and by allowing each dependent field to depend on both $\boldsymbol{\chi}_\bot$ and $\mathbf{x}_\bot$ and on both $\tau$ and $t$, in accord with the multiple scales asymptotic formalism.  By introducing these fast scales, we explicitly recognize the possibility for dynamics to occur on commensurate horizontal and vertical scales and on time scales of the order of the buoyancy period.  Such motions can be readily observed in visualizations from DNS of strongly stratified turbulence \citep[e.g. see][]{Rorai_2014,Waite_2014} and appear to be associated with various stratified shear instability mechanisms (e.g. Kelvin--Helmholtz billows and Holmboe waves). The fastest growing instability modes generally have streamwise wavelengths on the order of the shear layer thickness; i.e. in dimensional terms, $\mathit{O}(h)$. The DNS performed by \cite{Augier_2012,Augier_2015} provide further evidence of this scale separation and, in particular, of the importance of spectrally non-local energy transfers in stratified turbulence. {{Similarly, the recent DNS of \cite{Fritts_2021} provides direct evidence of the coupling between disparate-scale motions, with small-scale stratified turbulence generated by the interaction between breaking mountain waves significantly modifying large (horizontal) scale zonal flows.}}

Next, we identify a physically relevant and (what proves to be a) mathematically consistent distinguished limit; namely, the limit $Fr\to 0$ with $Pr$ fixed, $\alpha = Fr$ and the \textit{buoyancy} Reynolds number $Re_b\equiv Re Fr^2$ fixed.  Numerous DNS studies and scaling arguments  \citep{Smyth_2000,Billant_2001,brethouwer2007,Bartello_2013,Maffioli_2016} have isolated $Re_b$ as a key control parameter in stratified turbulence; here, it arises naturally when considering vertical diffusion of the large-scale buoyancy and horizontal momentum fields. {{Note that $Re_b$ is directly proportional to the Gibson number or `activity parameter' $Gn\equiv \varepsilon/(\nu N^2)$, often used in computational studies of stratified turbulence, when the turbulent scaling relation $\varepsilon = \tilde{c}U^3/L$ holds for some constant $\tilde{c}$. (See \cite{Portwood_2016} for further discussion.) Unlike the emergent parameter $Gn$, however, $Re_b$ is an external control parameter and thus more appropriate here.}} Once the reduced system has been derived, we are free to vary the numerical value of $Re_b$ to investigate its impact on flow transitions and various other flow features.

Owing to the scaling of the Reynolds stress feedbacks onto the fast-horizontal/fast-time mean flow, the expansion proceeds in fractional powers of $Fr$.  Accordingly, we introduce the asymptotic parameter $\epsilon\equiv\sqrt{Fr}$ and posit that
\begin{eqnarray}
[\mathbf{u}_\bot,b,p]&\sim&[\mathbf{u}_{0\bot},b_0,p_0]\,+\,\epsilon[\mathbf{u}_{1\bot},b_1,p_1]\,+\,\epsilon^2[\mathbf{u}_{2\bot},b_2,p_2]\,+\,\ldots,\label{UBPexpansions}\\
W&\sim&\frac{1}{\epsilon}W_{-1}\,+\,W_0\,+\,\epsilon W_1\,+\,\ldots\label{Wexpansion}
\end{eqnarray}
The expansions for $\mathbf{u}_\bot$, $b$ and $p$ start at $\mathit{O}(1)$, reflecting our expectation that the dominant contribution to each of these fields arises on large horizontal scales, in accord with our non-dimensionalization.  In contrast, in stratified turbulence, the vertical velocity is a \textit{small-scale} quantity~\citep{brethouwer2007,Maffioli_2016}.  Thus, in our two-scale formalism, we anticipate that $W$ will have a larger magnitude on the small horizontal scales ($\boldsymbol{\chi}_\bot$), where the flow will be shown to be non-hydrostatic, than on the large horizontal scales ($\mathbf{x}_\bot$).  Recalling that $Fr=\alpha$, we note that the dimensional velocity scale for $W$ simplifies to $\alpha U$ for the given distinguished limit, as expected on the basis of 3D incompressibility for flows with larger horizontal than vertical scales.  Thus, the rescaling of the dimensionless vertical velocity by the factor $1/\epsilon$ in (\ref{Wexpansion}) corresponds to re-nondimensionalizing the vertical velocity with $\sqrt{Fr}U$.  As shown subsequently, this rescaling simultaneously ensures that the fine scale dynamics occur on commensurate horizontal and vertical scales and, crucially, that the feedback of these ($\boldsymbol{\chi}_\bot$,$\tau$)-varying fluctuations onto the ($\mathbf{x}_\bot$,$t$)-varying mean fields arises at the proper order.  (Note that both the fluctuations and the mean fields vary on the same vertical coordinate $z$.)  This rescaling is also broadly consistent with the results of DNS showing that the long-time or ensemble \textit{mean-square} vertical velocity is $\mathit{O}(Fr)$, not $\mathit{O}(Fr^2)$, when normalized by $U^2$ \citep{Maffioli_2016}.

We proceed by substituting expansions (\ref{UBPexpansions}) and (\ref{Wexpansion}) into the Boussinesq equations (\ref{eqn:Bu})--(\ref{eqn:Bb}) with multiscale derivatives and collecting terms at like order in $\epsilon$.  We also introduce the fast-averaging operation $\overline{(\cdot)}$ for multiscale functions $\phi(\boldsymbol{\chi}_\bot,\mathbf{x}_\bot,z,\tau,t;\epsilon)$ defined such that
\begin{eqnarray}
\overline{\phi}(\mathbf{x}_\bot,z,t;\epsilon)&\equiv&\lim_{\tau_f,{l_x,l_y}\to\infty}\frac{1}{{l_x l_y} \tau_f}\int_0^{\tau_f}\int_\mathbf{D}\,\phi(\boldsymbol{\chi}_\bot,\mathbf{x}_\bot,z,\tau,t;\epsilon)\,\mathrm{d}\boldsymbol{\chi}_\bot \mathrm{d}\tau,\label{AVERAGING}
\end{eqnarray}
where $\mathbf{D}$ represents a horizontal $\boldsymbol{\chi}_{\bot}$-domain.  In the definition (\ref{AVERAGING}), the limiting process is associated with length scales ${l_x}$ and ${l_y}$ that are large compared to the scale of $\boldsymbol{\chi}_\bot$-variation of the function $\phi$ but small compared to the scale of $\mathbf{x}_\bot$-variation; $\tau_f$ can be interpreted analogously for the time integration.  In practice, we will take $\phi(\boldsymbol{\chi}_\bot,\mathbf{x}_\bot,z,\tau,t;\epsilon)$ to be doubly-periodic on $\mathbf{D}$, with `fast' spatial periods $l_x$ and $l_y$, so that the limits $l_x$, $l_y\to\infty$ need not be taken. As described \S~\ref{MTQL}, our novel methodology for integrating the reduced equations obviates the apparent need to specify the fast spatial periods ($l_x$, $l_y$) and time-integration period $\tau_f$ arbitrarily.
Finally, we note that the fast-averaging operation enables the multiscale fields to be decomposed into a slowly-varying mean component plus a fluctuation with zero fast-mean (denoted with a prime):  $\phi=\overline{\phi}+\phi'$.

At $\mathit{O}(\epsilon^{-2})$, we obtain the following system of equations:
\begin{eqnarray*}
\partial_\tau\mathbf{u}_{0\bot}\,+\,\left(\mathbf{u}_{0\bot}\cdot\nabla_{\boldsymbol{\chi}_{\bot}}\right)\mathbf{u}_{0\bot}&=&-\nabla_{\boldsymbol{\chi}_{\bot}}p_0,\\
\partial_z p_0 &=&b_0,\\
\partial_\tau b_0\,+\,\left(\mathbf{u}_{0\bot}\cdot\nabla_{\boldsymbol{\chi}_{\bot}}\right)b_0&=&0,\\
\nabla_{\boldsymbol{\chi}_{\bot}}\cdot\mathbf{u}_{0\bot}&=&0.
\end{eqnarray*}
From the $\mathit{O}(\epsilon^{-2})$ continuity equation, the leading-order horizontal velocity can be decomposed into a fast-horizontal ($\boldsymbol{\chi}_\bot$) average plus a vortical contribution that is non-divergent on the fast horizontal scales; i.e.
\begin{eqnarray*}
\mathbf{u}_{0\bot}&=&\overline{\mathbf{u}}^\chi_{0\bot}\,+\,\nabla_{\boldsymbol{\chi}}\times \Psi_0\hat{e}_z
\end{eqnarray*}
for scalar field $\Psi_0$, where $\overline{(\cdot)}^\chi$ indicates a fast-horizontal average, and $\hat{e}_z$ is a unit vector in the $z$ direction. Applying this fast horizontal averaging operation to the $\mathit{O}(\epsilon^{-2})$ horizontal momentum equation, it is readily shown that $\overline{\mathbf{u}}^\chi_{0\bot}=\overline{\mathbf{u}}_{0\bot}$; that is, the leading-order fast-horizontally-averaged horizontal velocity also must be independent of the fast time variable $\tau$. Substitution then yields
\begin{eqnarray*}
\partial_\tau\mathbf{u}^R_{0\bot}\,+\,\left(\overline{\mathbf{u}}_{0\bot}+\mathbf{u}^R_{0\bot}\right)\cdot\nabla_{\boldsymbol{\chi}_{\bot}}\mathbf{u}^R_{0\bot}&=&-\nabla_{\boldsymbol{\chi}_{\bot}}p_0,
\end{eqnarray*}
where $\mathbf{u}^R_{0\bot}\equiv \nabla_{\boldsymbol{\chi}}\times \Psi_0\hat{e}_z$. Crucially, there is no energy source for the purely vortical, fast-$\boldsymbol{\chi}_\bot$ non-divergent velocity field $\mathbf{u}^R_{0\bot}$, which would be strained by the mean flow $\overline{\mathbf{u}}_{0\bot}$ and ultimately dissipated on a time scale intermediate to $\tau$ and $t$ owing to shear-enhanced diffusion \citep{YoungJones_1991}. Accordingly, we henceforth set $\mathbf{u}^R_{0\bot}=\mathbf{0}$, so that
\begin{eqnarray}
\mathbf{u}_{0\bot}&=&\overline{\mathbf{u}}_{0\bot},
\end{eqnarray}
an important simplification in the subsequent analysis. The $\mathit{O}(\epsilon^{-2})$ horizontal momentum equation then requires $\nabla_{\boldsymbol{\chi}_{\bot}}p_0=0$, further implying from the vertical momentum equation that $\nabla_{\boldsymbol{\chi}_{\bot}}b_0=0$. The $\mathit{O}(\epsilon^{-2})$ buoyancy equation then gives
\begin{eqnarray}
b_0&=&\overline{b}_0,
\end{eqnarray}
which also is a source of considerable subsequent simplification. Finally, note that the fast average of the $\mathit{O}(\epsilon^{-2})$ vertical momentum equation yields
\begin{eqnarray}
\partial_z\overline{p}_0&=&\overline{b}_0.\label{eqn:STATICSm}
\end{eqnarray}
Thus, the large-scale flow is hydrostatically balanced at leading order.

Considering next the equations at $\mathit{O}(\epsilon^{-1})$, the fast-average of the continuity constraint 
\begin{eqnarray*}
\nabla_{\boldsymbol{\chi}_{\bot}}\cdot\mathbf{u}_{1\bot}\,+\,\partial_z W_{-1}&=&0
\end{eqnarray*}
gives $\partial_z\overline{W}_{-1}=0$, implying $\overline{W}_{-1}=0$. This deduction confirms that the vertical velocity is larger on the small scales than it is on the coarse scales (presuming $W'_{-1}\ne 0$).  Subtraction then gives the leading-order incompressibility condition for the fluctuating velocity field:
\begin{eqnarray}
\nabla_{\boldsymbol{\chi}_{\bot}}\cdot\mathbf{u}'_{1\bot}\,+\,\partial_z W'_{-1}&=&0.\label{eqn:CONTf}
\end{eqnarray}
At $\mathit{O}(\epsilon^{-1})$, an equation governing the leading-order dynamics of the fluctuating horizontal velocity field is obtained:
\begin{eqnarray}
\partial_\tau\mathbf{u}'_{1\bot}\,+\,\left(\overline{\mathbf{u}}_{0\bot}\cdot\nabla_{\boldsymbol{\chi}_{\bot}}\right)\mathbf{u}'_{1\bot}\,+\,W_{-1}'\partial_z\overline{\mathbf{u}}_{0\bot}\,+\,\nabla_{\boldsymbol{\chi}_{\bot}}p_1'&=&0.\label{eqn:perpMOMf}
\end{eqnarray}
In deriving this result we have tacitly assumed that the external forcing $\mathbf{f}_\bot$ is an $\mathit{O}(1)$ field. Similarly, after subtracting the fast-average result $\partial_z \overline{p}_1=\overline{b}_1$, the $\mathit{O}(\epsilon^{-1})$ fluctuation vertical momentum equation is 
\begin{eqnarray}
\partial_\tau W'_{-1}\,+\,\left(\overline{\mathbf{u}}_{0\bot}\cdot\nabla_{\boldsymbol{\chi}_{\bot}}\right)W'_{-1}\,+\,\partial_z p'_1-b'_1&=&0,\label{eqn:vertMOMf}
\end{eqnarray}
while at $\mathit{O}(\epsilon^{-1})$ the buoyancy equation reduces to 
\begin{eqnarray}
\partial_\tau b'_{1}\,+\,\left(\overline{\mathbf{u}}_{0\bot}\cdot\nabla_{\boldsymbol{\chi}_{\bot}}\right)b'_{1}\,+\,W_{-1}'\partial_z\overline{b}_{0}\,+\,W'_{-1}&=&0.\label{eqn:Bf}
\end{eqnarray}

To close the system, a set of constraints for the leading-order mean fields must be derived by fast-averaging the governing equations at $\mathit{O}(1)$. 
For example, the continuity equation for the slowly-varying mean fields is readily obtained:
\begin{eqnarray}
\nabla_{\mathbf{x}_{\bot}}\cdot\overline{\mathbf{u}}_{0\bot}\,+\,\partial_z \overline{W}_{0}&=&0.\label{eqn:CONTm}
\end{eqnarray}
Next, the $\mathit{O}(1)$ horizontal momentum equation can be expressed as
\begin{eqnarray*}
\partial_\tau\mathbf{u}'_{2\bot}\,+\,\left(\overline{\mathbf{u}}_{0\bot}\cdot\nabla_{\boldsymbol{\chi}_{\bot}}\right)\mathbf{u}'_{2\bot}&+&W'_{0}\partial_z\overline{\mathbf{u}}_{0\bot}\,+\,\nabla_{\boldsymbol{\chi}_{\bot}}p_2'\,=\,-\Bigg[\partial_t\overline{\mathbf{u}}_{0\bot}+\left(\overline{\mathbf{u}}_{0\bot}\cdot\nabla_{\mathbf{x}_{\bot}}\right)\overline{\mathbf{u}}_{0\bot}+\overline{W}_{0}\partial_z\overline{\mathbf{u}}_{0\bot}\\
&&\;+\,\left(\mathbf{u}_{1\bot}\cdot\nabla_{\boldsymbol{\chi}_{\bot}}\right)\mathbf{u}'_{1\bot}+W'_{-1}\partial_z\mathbf{u}_{1\bot}+\nabla_{\mathbf{x}_\bot}p_0-\frac{1}{Re_b}\partial_z^2\overline{\mathbf{u}}_{0\bot}-\mathbf{f}_{0\bot}\Bigg].
\end{eqnarray*}
For bounded behavior of the $\mathit{O}(\epsilon^2)$  fluctuation fields, a necessary condition is that the fast-average of the right-hand side of this equation must vanish. This solvability condition yields an equation for the slow evolution of the leading-order coarse-grained field $\overline{\mathbf{u}}_{0\bot}$:
\begin{eqnarray}
\partial_t\overline{\mathbf{u}}_{0\bot}+\left(\overline{\mathbf{u}}_{0\bot}\cdot\nabla_{\mathbf{x}_\bot}\right)\overline{\mathbf{u}}_{0\bot}+\overline{W}_0\partial_z\overline{\mathbf{u}}_{0\bot}&=&-\nabla_{\mathbf{x}_\bot}\overline{p}_0-\partial_z\left(\overline{W'_{-1}\mathbf{u}'_{1\bot}}\right)+\frac{1}{Re_b}\partial_z^2\overline{\mathbf{u}}_{0\bot}+\overline{\mathbf{f}}_{0\bot},\label{eqn:perpMOMm}
\end{eqnarray}
where (\ref{eqn:CONTf}), the incompressibility of the leading-order fluctuation velocity field, has been used.  The derivation of the equation governing the leading-order mean buoyancy field closely parallels that of the horizontal momentum equation. Specifically, at $\mathit{O}(1)$, 
\begin{eqnarray*}
\partial_\tau b_2'\,+\,\left(\overline{\mathbf{u}}_{0\bot}\cdot\nabla_{\boldsymbol{\chi}_{\bot}}\right)b_2'&+&W'_{0}\left(\partial_z\overline{b}_0+1\right)\,=\,-\Bigg[\partial_t\overline{b}_0+\left(\overline{\mathbf{u}}_{0\bot}\cdot\nabla_{\mathbf{x}_{\bot}}\right)\overline{b}_0+\overline{W}_{0}\left(\partial_z\overline{b}_0+1\right)\\
&&\qquad\qquad\qquad+\,\left(\mathbf{u}_{1\bot}\cdot\nabla_{\boldsymbol{\chi}_{\bot}}\right)b'_1+W'_{-1}\partial_z b_1-\frac{1}{Pr Re_b}\partial_z^2\overline{b}_0\Bigg].
\end{eqnarray*}
Fast-averaging this equation yields the evolution equation for $\overline{b}_0$:
\begin{eqnarray}
\partial_t\overline{b}_{0}+\left(\overline{\mathbf{u}}_{0\bot}\cdot\nabla_{\mathbf{x}_\bot}\right)\overline{b}_{0}+\overline{W}_0\partial_z\overline{b}_{0}+\overline{W}_0&=&-\partial_z\left(\overline{W'_{-1}b'_{1}}\right)+\frac{1}{Pr Re_b}\partial_z^2\overline{b}_{0}.\label{eqn:Bm}
\end{eqnarray}
Recalling (\ref{eqn:STATICSm}), the derivation of the leading-order mean and fluctuation equations is complete. Nevertheless, one further point regarding this derivation should be made. Inspection of the left-hand sides of the fluctuation equations following (\ref{eqn:CONTm}) and (\ref{eqn:perpMOMm}) along with the corresponding forms of the $\mathit{O}(1)$ fluctuation continuity and vertical momentum equations reveals that, for boundedness of the $\mathit{O}(\epsilon^2)$ fluctuation fields over the fast space and time scales, a \emph{second} solvability condition must be satisfied. Rather than yielding a slow evolution equation for the amplitude of the leading-order fluctuation fields, however, the required solvability condition simply constrains the evolution of the higher-order \emph{corrections} to the leading-order mean fields; see \cite{Michel_Chini_2019} for further details regarding this subtle but important point. Since these mean-field corrections are not required to close the leading-order mean/fluctuation system, we do not pursue the required calculation here.

\subsection{Synopsis of the reduced system}

Equations (\ref{eqn:perpMOMm}), (\ref{eqn:STATICSm}), (\ref{eqn:Bm}) and (\ref{eqn:CONTm}) and (\ref{eqn:perpMOMf}), (\ref{eqn:vertMOMf}), (\ref{eqn:Bf}) and (\ref{eqn:CONTf}) comprise a novel, multiscale reduced system.  For ease of reference this system is reproduced here, where, for brevity of notation, the numeric subscripts have been omitted and $\overline{\nabla}$ and $\nabla'$ are used in lieu of $\nabla_{\mathbf{x}_\bot}$ and $\nabla_{\boldsymbol{\chi}_\bot}$, respectively.\\

\noindent
\textbf{\underline{Mean Dynamics}}
\begin{eqnarray}
\left(\partial_t+\overline{\mathbf{u}}_\bot\cdot\overline{\nabla}+\overline{W}\partial_z\right)\overline{\mathbf{u}}_\bot
&=&-\overline{\nabla}\overline{p}\,-\,\partial_z\left(\overline{W'\mathbf{u}'_\bot}\right)\,+\,\frac{1}{Re_b}\partial_z^2\overline{\mathbf{u}}_\bot\,+\,\overline{\mathbf{f}}_\bot,\label{eqn:MEANuv}\\
0&=&-\partial_z\overline{p}\,+\,\overline{b},\label{eqn:MEANw}\\
\left(\partial_t+\overline{\mathbf{u}}_\bot\cdot\overline{\nabla}+\overline{W}\partial_z\right)\overline{b}
&=&-\overline{W}\,-\,\partial_z\left(\overline{W'b'}\right)\,+\,\frac{1}{Pr Re_b}\partial_z^2\overline{b},\label{eqn:MEANb}\\
\overline{\nabla}\cdot\overline{\mathbf{u}}_\bot\,+\,\partial_z\overline{W}&=&0.\label{eqn:MEANcont}
\end{eqnarray}
\noindent
\textbf{\underline{Fluctuation Dynamics}}
\begin{eqnarray}
\left(\partial_\tau+\overline{\mathbf{u}}_\bot\cdot\nabla'\right)\mathbf{u}'_\bot\,+\,W'\partial_z\overline{\mathbf{u}}_\bot
&=&-\nabla' p'\,+\,\frac{Fr}{Re_b}\left(\nabla'^2+\partial_z^2\right)\mathbf{u}'_\bot,\label{eqn:FLUCTuv}\\
\left(\partial_\tau+\overline{\mathbf{u}}_\bot\cdot\nabla'\right)W'&=&-\partial_z p'\,+\,b'\,+\,
\frac{Fr}{Re_b}\left(\nabla'^2+\partial_z^2\right)W',\label{eqn:FLUCTw}\\
\left(\partial_\tau+\overline{\mathbf{u}}_\bot\cdot\nabla'\right)b'\,+\,W'\partial_z\overline{b}
&=&-W'\,+\,\frac{Fr}{Pr Re_b}\left(\nabla'^2+\partial_z^2\right)b',\label{eqn:FLUCTb}\\
\nabla'\cdot\mathbf{u}'_\bot\,+\,\partial_z W'&=&0.\label{eqn:FLUCTcont}
\end{eqnarray}

\subsection{Generalized quasilinear (QL) structure of the reduced system}

Before attempting to integrate the multiscale reduced system numerically, it is instructive to examine the structure of these equations.   In the absence of the Reynolds-stress and buoyancy-flux divergence terms, the mean equations (\ref{eqn:MEANuv})--(\ref{eqn:MEANcont}) are readily seen to reduce to the hydrostatic primitive equations.  Heuristically, this set of equations governs the comparably slowly-evolving, large-scale layer-like motions that are routinely observed in the LAST regime of strongly stratified turbulence.  The correlations arising in these equations, which account for the collective effects of the small-scale flows on the strongly anisotropic large-scale motions, generally must be modeled phenomenologically.  By exploiting the scale separation associated with the chosen distinguished limit, however, here we are able to avoid the usual closure difficulties via explicit derivation of the leading-order equations governing the evolution of the fluctuation fields [i.e. (\ref{eqn:FLUCTuv})--(\ref{eqn:FLUCTcont})].  Taken together, equations (\ref{eqn:MEANuv})--(\ref{eqn:FLUCTcont}) comprise a \textit{closed} reduced system.  Direct calculation confirms that this system conserves energy in the absence of dissipation and forcing.

Inspection of (\ref{eqn:FLUCTuv})--(\ref{eqn:FLUCTb}) reveals that the fluctuations  are advected horizontally by the mean flow and can interact with vertical gradients of the mean buoyancy and horizontal velocity fields.  Equation (\ref{eqn:FLUCTw}) shows that the fluctuation dynamics are non-hydrostatic, as expected.  Moreover, the fluctuation dynamics are `quasilinear' with respect to the slowly-varying mean fields; that is, fluctuation/fluctuation nonlinearities are absent from the fluctuation equations, but their feedback on the coarse-grained fields is retained.  Indeed, a central outcome of the present work is that the increasingly popular QL approximation for anisotropic turbulent flows \citep{Fitzgerald_2014, constantinounavidetal2014, thomasetal2015, GallaireJFM2020, Fitzgerald_2018} and the associated statistical formulations (variously referred to as CE2 and SSST: \cite{tobiasdagonetal2011, srinivasanyoung2012, tobiasmarston2013, constantinoufarrelletal2014, aitchaaletal2015, Constantinou:2016fp, FarrellJFM2016, Farrell_Gayme_Ioannou_2016}) can be formally justified for strongly stratified (and perhaps other) shear flows via multiscale asymptotic analysis.  Since the mean fields are independent of the fast coordinates, the fluctuation equations are autonomous in $\boldsymbol{\chi}_\bot$ and there is no direct coupling among `fast' Fourier modes; instead, these modes are coupled only through their contribution to the modification of the mean fields.  By retaining slow spatiotemporal variability of the mean fields, our formulation in fact \textit{extends} the usual QL reduction:  precisely this sort of multiscale analysis provided inspiration for the so-called `generalized quasilinear' or GQL approximation~\citep{Marston:2016ff}, which has been demonstrated to improve the accuracy of QL-based predictions significantly for only a modest increase in model complexity \citep{Tobias:2016uq,chmt16}.  The reduced equations systematically derived here also can be compared to the 1D phenomenological model for fluctuating motions in stratified turbulence introduced in \cite{Rorai_2014} and extended by \cite{Feraco_2018}. In particular, we find that while the linear coupling and damping of the temperature and vertical velocity are adequately captured by the phenomenological model, neither the instability nor the nonlinear saturation terms in the fluctuation dynamics, which are fundamental to interactions with the large scales, are properly characterized in the 1D model.

A final point concerns the role of diffusive effects in the multiscale reduced system.  Owing to the anisotropic large-scale layering, vertical diffusion of momentum and buoyancy arises at leading-order in the mean equations.  Although asymptotically, i.e. as a function of $Fr$ as $Fr\to 0$, $Re_b=\mathit{O}(1)$ the effective diffusivity ($\propto 1/Re_b$) can be varied numerically to investigate different dynamical regimes captured by the reduced model.  Indeed, many DNS studies have suggested that important regime transitions occur at $Re_b\gtrsim 10$
\citep{Bartello_2013,Maffioli_2016,Lucas_2017_2,Lang_2019,Garanaik_2019}.  In contrast to the mean dynamics, the leading-order fluctuation equations are non-dissipative.  Nevertheless, in writing (\ref{eqn:FLUCTuv})--(\ref{eqn:FLUCTb}), we have included formally higher-order Laplacian diffusion terms as a simple means of regularizing the fluctuation dynamics, should large $z$-gradients (e.g.  possibly associated with critical layers) emerge. Although justifiable as a \textit{composite} asymptotic approximation, this approach reintroduces the small parameter $Fr$ into the limit system, and a more careful analysis would be required to ascertain whether other formally weak physical processes also may contribute to the dominant balance of terms in ultra-thin regions in which $z$-gradients become large.  
With this limitation understood, we proceed using the regularized reduced system (\ref{eqn:MEANuv})--(\ref{eqn:FLUCTcont}). Nonetheless, as shown explicitly in \S~\ref{RESULTS}, the 2D non-dissipative fluctuation equations can be re-expressed as the Taylor--Goldstein (TG) equation \citep{Craik_1985} since, formally, the mean fields are frozen during the fast evolution of the fluctuations.  Thus, by inspection, it is clear that the reduced system captures all of the linear instabilities admitted by the TG equation, including the classical Kelvin--Helmholtz and Holmboe instabilities.

\section{Integration of the reduced system}\label{ALGORITHM}

The structure of the reduced system (\ref{eqn:MEANuv})--(\ref{eqn:FLUCTcont}) is suggestive of a heterogeneous multiscale algorithm \citep{EngquistHMM2007}, in which fine-grained computations are performed on embedded domains in the local neighborhood of each coarse-scale grid point, thereby providing the fluxes that are needed to advance the coarse-scale fields.  An alternative approach would be to interpret (\ref{eqn:MEANuv})--(\ref{eqn:FLUCTcont})  as a physical space representation of the GQL formalism introduced by \cite{Marston:2016ff}.  Regardless, a multiscale implementation of some variety is required to exploit the full potential of the reduced model.  

In the present study, however, we adopt the more modest goal of illustrating the dynamics that can be exhibited by the reduced system in small 2D domains {{with commensurate horizontal and vertical lengths (each of order $h$ in dimensional terms)}}. To this end we suppress the slow $\mathbf{x}_\bot$ derivatives, which immediately implies that $\overline{W}=0$.  The resulting (2D) reduced system can be expressed as
\begin{eqnarray}
\partial_t \overline{u}&=&{\partial_z\left(\overline{\partial_z \psi' \partial_\chi\psi'}\right)}\,+\,\frac{1}{Re_b}\partial_z^2\overline{u}+\overline{{f}},\label{QL_u_bar}\\
\partial_t\overline{b}&=&{\partial_z\left(\overline{b'\partial_\chi\psi'}\right)}\,+\,\frac{1}{Pr Re_b}\partial_z^2\overline{b},\label{QL_b_bar}\\
(\partial_\tau + \overline{u} \partial_\chi) \triangle \psi' &=&\partial_\chi\psi'\partial_z^2\overline{u}-\partial_\chi b'\,+\,\frac{{Fr}}{Re_b}\triangle^2 \psi',\label{QL_up}\\
(\partial_\tau + \overline{u} \partial_\chi) b'&=&(1 + \partial_z\overline{b}) \partial_\chi \psi'\,+\,\frac{{Fr}}{Pr Re_b}\triangle b'\label{QL_bp}
\end{eqnarray}
where $u'\equiv\partial_z\psi'$, $W'\equiv -\partial_\chi\psi'$ and $\triangle = \partial_\chi^2+\partial_z^2$. Although the dependence on the slow spatial coordinate(s) $\mathbf{x}_\bot$ has been suppressed, the reduced system (\ref{QL_u_bar})--(\ref{QL_bp}) nevertheless still formally requires time advancement on \textit{two} time scales ($\tau$ and $t$).  Below, we employ two distinct strategies for integrating the reduced system. The first treats (\ref{QL_u_bar})--(\ref{QL_bp}) as an initial-value problem on the fast time scale. The second is based on a new asymptotic analysis of slow-fast QL systems with fast instabilities and exploits the tendency of these systems to self-tune to a state of approximate marginal stability \citep{Michel_Chini_2019}.

\subsection{Single time-scale formulation \label{STQL}}

One straightforward way to simulate the reduced equations \eqref{QL_u_bar}--\eqref{QL_bp} is to revert to a single time-scale formulation by making the non-asymptotic replacement $\partial_t=(1/Fr)\partial_\tau$ and by reinterpreting the fast ($\boldsymbol{\chi}_\bot$,$\tau$) average as a strict horizontal average (only).
This reformulation sacrifices certain advantages accrued via the multiples scales asymptotic analysis. In particular, the resulting system is numerically stiff owing to the reintroduction of $Fr$, and both the fast fluctuations and the slowly evolving mean fields have to be numerically co-evolved on the fast time scale $\tau$. The advantage of this approach is that there is a clear protocol for discretizing this system in both space and time. In fact, the resulting equations are identical to those that would be obtained via an \textit{ad hoc} QL approximation of the (2D) Boussinesq equations, in which flow fields are decomposed into a horizontal (`streamwise') average plus a fluctuation about that mean \citep{Fitzgerald_2018,Fitzgerald_2019}.  Regarding the spatial discretization, owing to the QL structure \emph{any} set of horizontal Fourier modes can be included. Nevertheless, since this set must be specified \emph{a priori}, standard grids equispaced in physical or Fourier space generally are employed. As described in the following subsection, a chief virtue of the multiple time-scale formulation is that the intrinsic dynamics of the slow--fast QL system self-selects those wavenumbers and associated Fourier modes to be included. Consequently, in that formulation, the fast spatial domain is effectively infinite in extent; there is no \emph{a priori} quantization of Fourier modes imposed by the seemingly arbitrary specification of $l_x$.

\subsection{Multiple time-scale formulation \label{MTQL}}

In our view, an algorithm that explicitly enforces the time scale separation between the mean and fluctuation fields in \eqref{QL_u_bar}--\eqref{QL_bp} is preferable. The method for integrating slow--fast QL systems with fast instabilities introduced by \cite{Michel_Chini_2019} leverages the linearity and autonomy of  the fluctuation dynamics on the fast time scale, which here implies that both $\psi'$ and $b'$ depend on the fast time $\tau$ only through a term of the form $e^{\sigma \tau}$, where $\sigma$ is the (complex-valued, linear) growth rate. For the Reynolds stress divergence to remain finite in the mean field equations \eqref{QL_u_bar} and \eqref{QL_b_bar}, either the real part of the growth rate or the amplitude of the fluctuations therefore must vanish. The crucial point, as we demonstrate below, is that for forced strongly stratified flows the fluctuation amplitude (if non-zero) is then slaved to the mean fields to ensure that the real part of the growth rate $\sigma_r = 0$; that is, to ensure that the Reynolds stress divergence modifies the evolution of $\bar{u}$ and $\bar{b}$ so that the mean fields always evolve (slowly) on a marginal-stability manifold. This constrained evolution is a mathematical manifestation of self-organized criticality, a physical attribute increasingly being associated with stratified turbulent shear flows \citep{SalehipourJFM2018,SmythNature2019}, although the argument that stratified flows adjust to a marginally stable `pseudo-equilibrium' dates back at least to \cite{Turner_1973}. The remainder of this section describes in detail the resulting multiscale algorithm.

First, note that fluctuation system \eqref{QL_up}-\eqref{QL_bp} can be treated as a linear homogeneous eigenvalue problem by seeking modal solutions with real-valued wavenumber $k(t)$, (complex-valued) growth rate $\sigma(\overline{\mathbf{G}},k)$, where $\overline{\mathbf{G}}$ denotes dependencies on the mean fields and their derivatives (i.e. $\bar{u}$, $\partial_z^2 \bar{u}$ and $\partial_z \bar{b}$),  and complex-valued amplitude $A$ of slowly-varying magnitude $\vert A(t)\vert$:
\begin{align}
\psi'(\chi, z, \tau; t) &= A(t) \hat{\Psi}(z;t) e^{\sigma \tau + i k(t) \chi} + \mathrm{c.c.},\\
b'(\chi, z, \tau; t) &= A(t) \hat{b}(z;t) e^{\sigma \tau + i k(t) \chi} + \mathrm{c.c.},
\end{align}
where {{$\sigma=\sigma_r+i\sigma_i$ for real $\sigma_r$ and $\sigma_i$}},
$\mathrm{c.c.}$ denotes complex conjugate and we have indicated explicitly that the vertical eigenfunctions $\hat{\Psi}$ and $\hat{b}$ also may vary slowly with time.  Recalling that the fluctuation horizontal velocity and buoyancy are $\mathit{O}(\epsilon)$ relative to the corresponding mean fields, it is clear from these expressions that the uniformity of the posited asymptotic expansions in \eqref{UBPexpansions} would be lost if $\sigma_r>0$. Indeed, in that situation, the fluctuations would grow exponentially fast -- on the fast time scale -- while the mean fields remained locally frozen in time. Accordingly, as an asymptotic uniformity condition, we seek a solvability constraint that ensures $\sigma_r\le 0$. In practice, there may be `instants' during the slow evolution where this condition cannot be satisfied, in which case the fluctuations will attain large (but finite) amplitude and the $\mathit{O}(1)$ mean fields will respond rapidly; i.e. on the fast time scale. In these situations, temporal scale separation is \textit{transiently} lost. Although the algorithm we develop can be modified to incorporate this intermittent bursting behavior properly \citep{Ferraro_2019}, this extension proves unnnecessary for the parameter regime explored in the present investigation and is not described here. In contrast, $\sigma_r<0$ presents no conceptual or pragmatic difficulty, as disturbances are exponentially damped on the fast time scale and, therefore, the amplitude $A$ is self-consistently set to zero.
If $\sigma_r=0$, then the averaging required for evaluation of the Reynolds stress divergence terms in the mean-field equations [e.g. $\partial_z \left(\overline{\partial_z \psi' \partial_\chi\psi'}\right)$ in \eqref{QL_u_bar}] is well-defined. Specifically,
\begin{align}
\partial_z\left(\overline{\partial_z \psi' \partial_\chi\psi'}\right)&= {\vert A(t) \vert}^2 ik \partial_z \left( \hat{\Psi} \partial_z \hat{\Psi}^* - \hat{\Psi}^* \partial_z \hat{\Psi} \right) \equiv  {\vert A(t) \vert}^2 \mathrm{RS}_u,\\
\partial_z\left(\overline{b'\partial_\chi\psi'}\right)&=  {\vert A(t) \vert}^2 ik \partial_z \left( \hat{\Psi} \hat{b}^* - \hat{\Psi}^* \hat{b} \right) \equiv  {\vert A(t) \vert}^2 \mathrm{RS}_b,
\end{align}
where $\mathrm{RS}_u$ and $\mathrm{RS}_b$, defined above, have been introduced for brevity of notation, and an asterisk denotes complex conjugation. Thus, the QL system \eqref{QL_u_bar}-\eqref{QL_bp} reduces to the hybrid slow-initial-value/fast-eigenvalue problem:
\begin{eqnarray}
\partial_t \overline{u}&=& {\vert A(t) \vert}^2 \mathrm{RS}_u\,+\,\frac{1}{Re_b}\partial_z^2\overline{u}+\overline{{f}},\label{MT_lin_sys_4}\\
\partial_t\overline{b}&=& {\vert A(t) \vert}^2 \mathrm{RS}_b\,+\,\frac{1}{Pr Re_b}\partial_z^2\overline{b},\label{MT_lin_sys_5}\\
(\sigma + ik \overline{u} )(\partial_z^2 - k^2) \hat{\Psi} &=&ik(\partial_z^2\overline{u}) \hat{\Psi}-ik \hat{b}\,+\,\frac{{Fr}}{Re_b}(\partial_z^2 -k^2)^2 \hat{\Psi},\label{MT_lin_sys_1}\\
(\sigma +ik  \overline{u}) \hat{b}&=&ik(1 + \partial_z\overline{b})\hat{\Psi}\,+\,\frac{{Fr}}{Pr Re_b}(\partial_z^2 -k^2)\hat{b},\label{MT_lin_sys_2}
\end{eqnarray}
where ${\vert A(t) \vert}=0$ if $\sigma_r < 0$.  

The central challenge is to determine self-consistently an equation for the \textit{a priori} unknown amplitude function ${\vert A(t) \vert}$, which, of course, is not constrained via the solution of the linear homogeneous eigensystem \eqref{MT_lin_sys_1}--\eqref{MT_lin_sys_2}. As demonstrated in \cite{Michel_Chini_2019}, to derive this constraint on ${\vert A(t) \vert}$ in the case of marginal stability, i.e. when $\sigma_r=0$, the linear eigensystem itself must be differentiated with respect to the slow time variable and an appropriate solvability condition enforced. This slow-time differentiation enables nonlinearity present in the coupled system \eqref{MT_lin_sys_4}--\eqref{MT_lin_sys_2}  to be incorporated, as required for determination of the fluctuation amplitude. 

To facilitate the required analysis, the linear subsystem \eqref{MT_lin_sys_1}--\eqref{MT_lin_sys_2} first is recast in the form $\mathcal{L}X = 0$, with $X = [\hat{\Psi}(z),\hat{b}(z)]^T$ and
\begin{equation}
\mathcal{L} = \begin{pmatrix} (\sigma +ik \overline{u})(\partial_z^2-k^2) -ik \partial_z^2 \overline{u}- \frac{Fr}{Re_b}(\partial_z^2-k^2)^2&ik\\-ik(1+\partial_z \overline{b})& \sigma+ik\overline{u}- \frac{Fr}{Pr Re_b}(\partial_z -k^2) \end{pmatrix}, 
\end{equation} 
complemented with periodic boundary conditions in $z$. With respect to the eigenproduct
\begin{equation}
(X_1 \vert X_2) \equiv \int_0^{l_z} X_1(z) X_2^*(z) \mathrm{d}z~~~~~\forall~(X_1,X_2),
\end{equation}
the adjoint operator of $\mathcal{L}$, defined via $(\mathcal{L}X_1\vert X_2) = (X_1\vert \mathcal{L}^\dagger X_2)$, is given by
\begin{equation}
\mathcal{L}^\dagger = \begin{pmatrix} (\sigma^* -ik \overline{u})(\partial_z^2-k^2) -2ik \partial_z \overline{u} \partial_z- \frac{Fr}{Re_b}(\partial_z^2-k^2)^2&ik(1+\partial_z \overline{b})\\-ik& \sigma^*-ik\overline{u}- \frac{Fr}{Pr Re_b}(\partial_z -k^2) \end{pmatrix}.
\end{equation} 
Note that $\mathcal{L}$ is not self-adjoint ($\mathcal{L}\neq \mathcal{L}^\dagger$). The linear problem $\mathcal{L}X = 0$ is then differentiated with respect to the slow time $t$, yielding 
\begin{equation}
\mathcal{L} \frac{\mathrm{d} X}{\mathrm{d} t}\,=\,-\frac{\mathrm{d} \mathcal{L}}{\mathrm{d} t} X,\label{dXdtEQN}
\end{equation}
where
\begin{equation}
\frac{\mathrm{d} \mathcal{L}}{\mathrm{d} t}= \begin{pmatrix} (\mathrm{d}\sigma / \mathrm{d}t + ik \partial_t \overline{u}) (\partial_z^2-k^2)-ik (\partial_z^2 \partial_t \overline{u})& 0\\-ik \partial_z\partial_t\overline{b}& \mathrm{d}\sigma/\mathrm{d}t +ik\partial_t \overline{u}\end{pmatrix} {+ \frac{\mathrm{d}k}{\mathrm{d}t} M}
\end{equation}
and $M$ is a matrix whose explicit computation turns out not to be necessary. 
Since $\mathcal{L}$ is singular, 
the linear system \eqref{dXdtEQN} is solvable if and only if the right-hand side is orthogonal to the corresponding null adjoint eigenvector $X^\dagger$ (the Fredholm alternative condition). This requirement can be readily confirmed by forming the inner product of left-hand side of \eqref{dXdtEQN} with respect to $X^\dagger$, i.e.
\begin{equation}
\left. \left( \mathcal{L} \frac{\mathrm{d} X}{\mathrm{d} t} \right\vert X^\dagger \right) = \left.\left(\frac{\mathrm{d} X}{\mathrm{d} t} \right\vert \mathcal{L}^\dagger X^\dagger \right) = \left.\left( \frac{\mathrm{d} X}{\mathrm{d} t} \right\vert 0 \right) = 0. 
\end{equation}
Consequently, using \eqref{dXdtEQN}, we obtain
\begin{equation}
\left.\left( \frac{\mathrm{d} \mathcal{L}}{\mathrm{d} t} X \right\vert X^\dagger \right)=0.\label{MT_lin_sys_3}
\end{equation}
Crucially, this constraint can be made explicit.  Using the mean-field equations \eqref{MT_lin_sys_4} and \eqref{MT_lin_sys_5} and noting that $X^\dagger=[\hat{\Psi}^{\dagger}(z), \hat{b}^{\dagger}(z)]^T$, the solvability condition becomes
\begin{equation}
C_1 \frac{\mathrm{d}\sigma}{\mathrm{d}t} = C_2  {\vert A(t) \vert}^2\,+\,C_3\,+\,C_4 \frac{\mathrm{d}k}{\mathrm{d}t}.\label{sigma_solvability}
\end{equation}
As for the matrix $M$, the computation of the coefficient $C_4$ proves unnecessary. The remaining coefficients are given by the following expressions:
\begin{align}
C_1 &= \frac{1}{ik}\int_0^{l_z}\left[\hat{\Psi}^{\dagger *} (\partial_z^2 -k^2) \hat{\Psi} + \hat{b}^{\dagger *} \hat{b} \right] \mathrm{d}z,\\
C_2 &= \int_0^{l_z}\left\lbrace \mathrm{RS}_u \left[\hat{\Psi}(\partial_z^2 +k^2)    \hat{\Psi}^{\dagger *}  + 2 \partial_z \hat{\Psi} \partial_z    \hat{\Psi}^{\dagger *}  - \hat{b}\hat{b}^{\dagger *}  \right] - \mathrm{RS}_b \left[\hat{b}^{\dagger *} \partial_z \hat{\Psi} + \hat{\Psi} \partial_z \hat{b}^{\dagger *} \right]  \right\rbrace \mathrm{d}z,\\
C_3 &= \int_0^{l_z}\left\lbrace  \left( \overline{f} + \frac{\partial_z^2 \overline{u}}{Re_b} \right) \left[\hat{\Psi}(\partial_z^2 + k^2)  \hat{\Psi}^{\dagger *} + 2 \partial_z \hat{\Psi} \partial_z  \hat{\Psi}^{\dagger *} - \hat{b} \hat{b}^{\dagger *}  \right] - \frac{\partial_z^2 \overline{b}}{Pr Re_b} \left( \hat{b}^{\dagger *} \partial_z \hat{\Psi} + \hat{\Psi} \partial_z \hat{b}^{\dagger *} \right)\right\rbrace \mathrm{d}z.
\end{align}
Dividing both sides of (\ref{sigma_solvability}) by $C_1$, the evolution of the linear growth rate $\sigma_r$ therefore is given for \emph{fixed} $k$ by 
\begin{equation}
{ \left( \frac{\partial\sigma_r}{\partial t} \right)_{k} }= \alpha_r - \beta_r  {\vert A(t) \vert}^2,\label{dsigmadt}
\end{equation}
where both $\alpha_r = \mathrm{Re}\{C_3/C_1\}$ and $\beta_r = \mathrm{Re}\{-C_2/C_1\}$ are computable functionals of the slow fields $\overline{u}$ and $\overline{b}$, the forcing $\overline{f}$, the direct and adjoint eigenfunctions $[\hat{\Psi}, \hat{b}]^T$ and $[\hat{\Psi}^\dagger, \hat{b}^\dagger]^T$, respectively, the parameters $Re_b$ and $Pr$ and the wavenumber $k$. Note that the total temporal variation of the real part of the growth rate $\sigma(\overline{\mathbf{G}},k)$,
\begin{equation}
\frac{\mathrm{d}\sigma_r}{\mathrm{d}t} =  \left( \frac{\partial\sigma_r}{\partial t} \right)_{k}  + \frac{\mathrm{d}k}{\mathrm{d}t}  \left(\frac{\partial\sigma_r}{\partial k} \right)_{\overline{\mathbf{G}}},
\end{equation}
includes an additional contribution owing to the time variation of $k$. The second term on the right-hand side of this expression vanishes, however, provided that $k(t)$ is (locally) the most unstable mode, as required here. As discussed in detail in \cite{Michel_Chini_2019}, temporal scale separation then requires (cf. (\ref{dsigmadt})) that
\begin{equation}
 {\vert A(t) \vert} = \Bigg\{ 
\begin{array}{l l}
    \sqrt{\alpha_r / \beta_r} & \quad \text{if}~ \sigma_r=0, ~\alpha_r >0 ~\mathrm{and}~ \beta_r>0;\\
  0 & \quad \text{otherwise}.\\ 
\end{array} 
\label{A_mt}
\end{equation}
In contrast to its magnitude, the phase of $A(t)$ remains unconstrained in this formalism and, in fact, is not needed to evolve the reduced dynamics. Indeed, the reduced system \eqref{MT_lin_sys_4}-\eqref{MT_lin_sys_2} is invariant under any such phase change. In the fully nonlinear Boussinesq equations, this quantity would be fixed by the spatial phases of those fluctuation modes of generally small initial amplitude that become linearly unstable on the fast time scale; that is, this phase information ultimately depends on the details of the initial conditions, which are filtered in this framework. A primary virtue of this formalism is precisely that numerical time-integration need only be performed on the \textit{slow} time scale $t$ without the arbitrary reinitialization of the fluctuation fields at each slow time instant.
Specifically, at each slow time step, the following algorithm is executed.
\begin{enumerate}
\item The direct eigenvalue problem $\mathcal{L}X=0$ is solved with periodic boundary conditions in $z$, and both the eigenvalue $\sigma$ of largest real part and the corresponding eigenvector $X$ are computed.
\item Step (i) is repeated over adjacent $k$ (i.e.~$k- \Delta k$ and $k+\Delta k$, for suitably small horizontal wavenumber increment $\Delta k$) to reach a wavenumber $k$ corresponding to a local maximum of $\sigma_r(k)$.
\item If $\sigma_r(k)=0$ (numerically, if $\sigma_r(k)\geqslant 0$), the adjoint eigenvalue problem $\mathcal{L}^\dagger X^\dagger = 0$ is solved with periodic boundary conditions in $z$, and a nonzero eigenvector $X^\dagger$ corresponding to the null eigenvalue is returned. The modal  amplitude ${\vert A(t) \vert}$ is then computed through \eqref{A_mt}.
\item The mean fields $\overline{u}$ and $\overline{b}$ are then time-advanced using a suitable discretization of \eqref{MT_lin_sys_4}--\eqref{MT_lin_sys_5}.
\end{enumerate} 

Finally, note that in the present formulation, $k$ is determined by scanning over a range of wavenumbers at each slow time (step~(ii) of the algorithm detailed above) to identify a local maximum of $\sigma_r(k)$. In companion work, we are seeking an extension of this algorithm that obviates the need to solve the eigenvalue problem repeatedly for different wavenumbers $k$ by deriving a set of slow equations for both ${\vert A(t) \vert}$ \textit{and} $\mathrm{d}k/\mathrm{d}t$; see \cite{Ferraro_2019}. Regardless of the algorithmic details, the wavenumber $k$ in the multiple time-scale formulation can smoothly vary on the slow time scale $t$. Crucially, this variation makes possible the accurate computation of the wavenumber that would be realized in the limit of an \textit{infinite} horizontal domain, thereby ameliorating a recurring issue in the study of, e.g., the nonlinear dynamics of parallel shear flows and transition to turbulence \citep{Tuckerman_2020}.

\section{Illustrative application to strongly stratified Kolmogorov flow}\label{RESULTS}

In this section, we explore the dynamics of the reduced equations (\ref{QL_u_bar})--(\ref{QL_bp}) when the flow is driven by an imposed (deterministic) mean body force that contains a single vertical wavenumber $m$: $\overline{f}=(m^2/Re_b)\cos(mz)$.  The coefficient $m^2/Re_b$ is chosen so that the resulting steady laminar velocity profile that would be driven in the absence of instabilities is simply $\overline{u}_L(z) = \cos(mz)$ (where a subscript `$L$' is used to refer to the laminar or base state); i.e. we consider (2D) stratified Kolmogorov flow.  For specificity, we fix the values of the parameters $m=3$ and $Pr=1$.
Note that by choosing an order unity numerical value for the forcing wavenumber $m$, we are directly driving `velocity layers,' each having a dimensional thickness of order $h=U/N$. To facilitate careful assessment of the performance of our multiple time-scale algorithm, we restrict the vertical extent of our computational domain to encompass only a single pair of shear layers, recognizing that in so doing we preclude the occurrence of potentially important long-wavelength instabilities in the vertical direction; in particular, see \cite{Balmforth_2002}, \cite{Balmforth_2005} and \cite{Garaud_2015}.

As documented in \S~\ref{LSA}, linear stability analysis confirms that the base flow $\bar{u}_L$ is strongly unstable to small-amplitude perturbations. The resulting instabilities excite vertical motions that enhance mixing and sustain a buoyancy staircase. Consequently, the chosen stratified Kolmogorov flow configuration provides a framework to perform both (i)~a quantitative comparison between DNS of the governing 2D Boussinesq equations and the reduced model derived for the LAST regime in the limit of small Froude and large Reynolds number, and (ii)~an investigation of the mixing enhancement and mean buoyancy profile as $Re_b$ is increased.

\subsection{Linear stability analysis}\label{LSA}

We begin by considering the linear stability of $\overline{u}_L(z) = \cos(3z)$ in the presence of the imposed linear stratification (and zero mean perturbation, so that $\overline{b}_L(z)=0$), as governed by the subsystem (\ref{QL_up})--(\ref{QL_bp}).  Making a normal mode ansatz for a disturbance mode with $\chi$-wavenumber $k$, $\psi'(\chi,z,\tau)=\hat{\psi}(z)\mbox{e}^{i k(\chi-c\tau)}+\mbox{c.c.}$, where the complex phase speed $c\equiv \omega/k$ (for angular frequency $\omega$), yields the TG equation in the non-dissipative limit $Fr/Re_b\to 0$:
\begin{eqnarray}
\frac{d^2\hat{\psi}}{dz^2}\,+\,\left[\frac{\left(1+\partial_z\overline{b}_L\right)}{\left(\overline{u}_L-c\right)^2}-\frac{\partial_z^2\overline{u}_L}{\left(\overline{u}_L-c\right)}-k^2\right]\hat{\psi}&=&0.\label{eqn:TG}
\end{eqnarray}
The local-in-time gradient Richardson number $Ri_g=(1+\partial_z\overline{b})/|\partial_z\overline{u}|^2$.  For the given basic-state profile,
$\overline{u}_L(z) = \cos(3z)$ and $\overline{b}_L(z)=0$, so the gradient Richardson number of the laminar state
\begin{eqnarray}
Ri_{gL}&=&\frac{1}{m^2\sin^2(mz)},\label{Rigl}
\end{eqnarray}
which attains a minimum value $Ri_{gL_{\mathrm{min}}}=1/m^2$ for $z=\pm\pi/(2m),\pm 3\pi/(2m) \ldots$  Thus, for the chosen parameter values $Ri_{gL_{\mathrm{min}}}=1/9<1/4$, so the Miles--Howard necessary criterion for linear instability is satisfied.  Of course, the incorporation of diffusion will modify this inviscid criterion, but the effective diffusivity $Fr/Re_b$ is much less than unity in the subsequent numerical simulations and acts only as a regular perturbation to the non-dissipative problem in smooth regions of the flow. 

\subsection{Nonlinear evolution\label{Nonlinear_evol}}

To assess the performance of the reduced system quantitatively, a set of numerical simulations is performed for $Re_b=1$ and decreasing values of $Fr$ down to $10^{-2}$. Three different numerical algorithms are compared, each implemented in the computing environment \textit{Dedalus} \citep{Dedalus_2020}. In each case, a pseudo-spectral method is used with a second-order Runge-Kutta time-stepping scheme. Python codes are provided as electronic {{supplementary material}}.

First, a set of direct numerical simulations (DNS) of the primitive 2D Boussinesq equations, that is \eqref{eqn:Bu}--\eqref{eqn:Bb} rendered two-dimensional, is performed. With $\alpha=Fr$, we note that the dimensionless total energy density is $(u^2+Fr^2 w^2+b^2)/2$. The domain, of vertical size $l_z=2\pi/3$ and of horizontal size $L_x=(2\pi/k)Fr$, is discretized using a Fourier--Fourier pseudo-spectral method with 128 grid points in each direction. Note that $k$ must be specified initially and remains fixed for the duration of the simulation. 

The second algorithm is the single time-scale formulation of the QL system (STQL) introduced in \S~\ref{STQL}. These simulations also are set in a domain  of size $l_z=2\pi/3$ and $l_x=2\pi/k$ using $128$ grid points (cf. Fourier modes) in each direction. As for the DNS, $k$ must be fixed \textit{a priori}. Unlike the DNS, certain terms in the governing Boussinesq equations are consistently neglected in the STQL simulations, in accord with the multiple scales analysis performed in the limit $Fr \rightarrow 0$ with $Re_b$ fixed.

The third numerical scheme integrates the multiple time-scale formulation of the QL system (MTQL). The equations governing the 1D evolution of the slowly-evolving mean fields $\bar{u}(z,t)$ and $\bar{b}(z,t)$, \eqref{QL_u_bar} and \eqref{QL_b_bar}, respectively, are discretized using Chebyshev polynomials -- rather than Fourier modes, for compatibility with the \textit{Dedalus} eigenvalue solver -- with 128 grid points. As detailed in \S~\ref{MTQL}, the filtered dynamics of the fast fluctuations is obtained by solving at each slow time-step a 1D eigenvalue problem, yielding the instantaneous linear growth rate $\sigma_r$ and vertical mode structure ($\hat{\Psi}(z;t)$ and $\hat{b}(z;t)$) of the most unstable mode. Once the wavenumber $k$ is updated and determined to correspond to a local maximum of $\sigma_r(k)$ within $\Delta k=10^{-3}$, the amplitude of the fluctuation mode is set in accord with condition \eqref{A_mt}. Again, we emphasize that in the MTQL algorithm the wavenumber $k$ is \textit{not} fixed, but evolves so that the marginally stable wavenumber is tracked, with modes corresponding to adjacent values of $k$ being linearly stable. In principle, the amplitude $A(t)$ of the fluctuation mode would be updated continually such that $\sigma_r = 0$ remains fixed; in practice, $\sigma_r$ would remain very close to the first positive value computed, which depends on the specific value of the time step $\Delta t$ but, in any case, will be very small compared to unity (e.g. $\sigma_r = 2\times 10^{-4}$ for $Fr=0.02$, $Re_b=1$ and $\Delta t=10^{-4}$). To achieve a steady state in which $\sigma_r$ is independent of the time step, the modal amplitude $A(t)$ is artificially increased during the transient regime such that
\begin{equation}
\textrm{if } \sigma_r > 10^{-6} \textrm{ and } A>0, \textrm{ then } A^2\,  =\,\frac{\alpha_r}{\beta_r} + \frac{\sigma_r}{1000 \beta_r {\Delta t}}. 
\end{equation}
According to \eqref{dsigmadt}, $\sigma_r$ will, if $A$ is subject to this change, vary during the ($n+1)$-st time step according to $\Delta \sigma_r\equiv \sigma_r^{n+1}-\sigma_r^n=-\sigma_r^{n}/1000$, and therefore will relax exponentially to zero provided the stated conditions are satisfied. Note that this modification minimally affects the transient regime, and is no longer applied once a steady-state is reached, since then $\sigma_r < 10^{-6}$.

The same time step $\Delta t = 10^{-4}$ is used in all of the numerical simulations.  It should be emphasized, however, that because the MTQL system is evolved on the slow time scale much larger values of $\Delta t$ could be utilized for the MTQL simulation. Although a detailed study of the computational savings afforded by the MTQL algorithm is beyond the scope of this investigation, we note here that the steady state depicted subsequently for $Fr =0.02$ and $Re_b=1$ is also quantitatively reproduced with the MTQL algorithm using a time step $\Delta t = 0.01$ (i.e. 100 times larger) for which the DNS algorithm is numerically unstable.

\begin{figure}
\begin{center}
\includegraphics[width=0.49\linewidth]{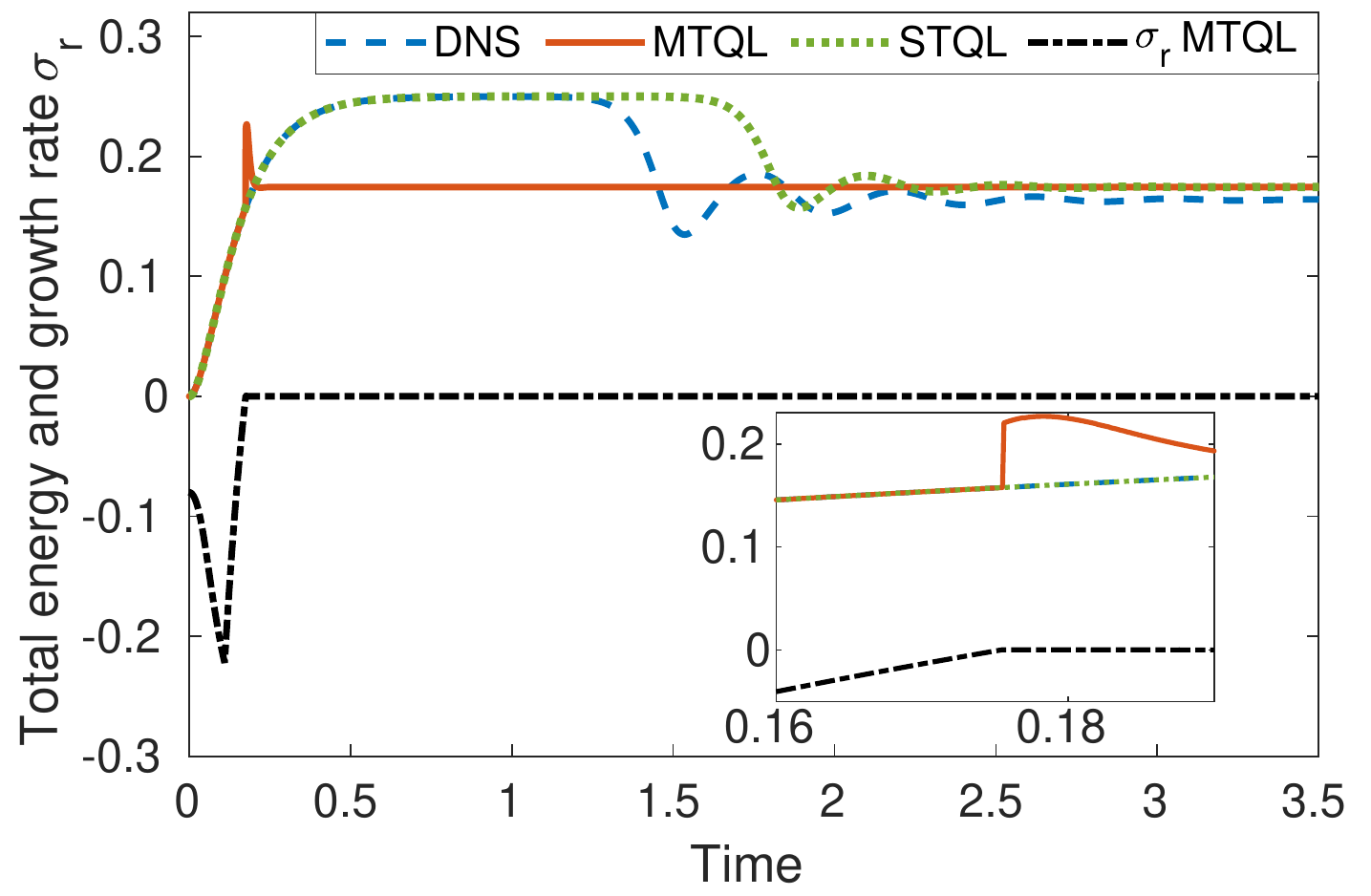}\includegraphics[width=0.55\linewidth]{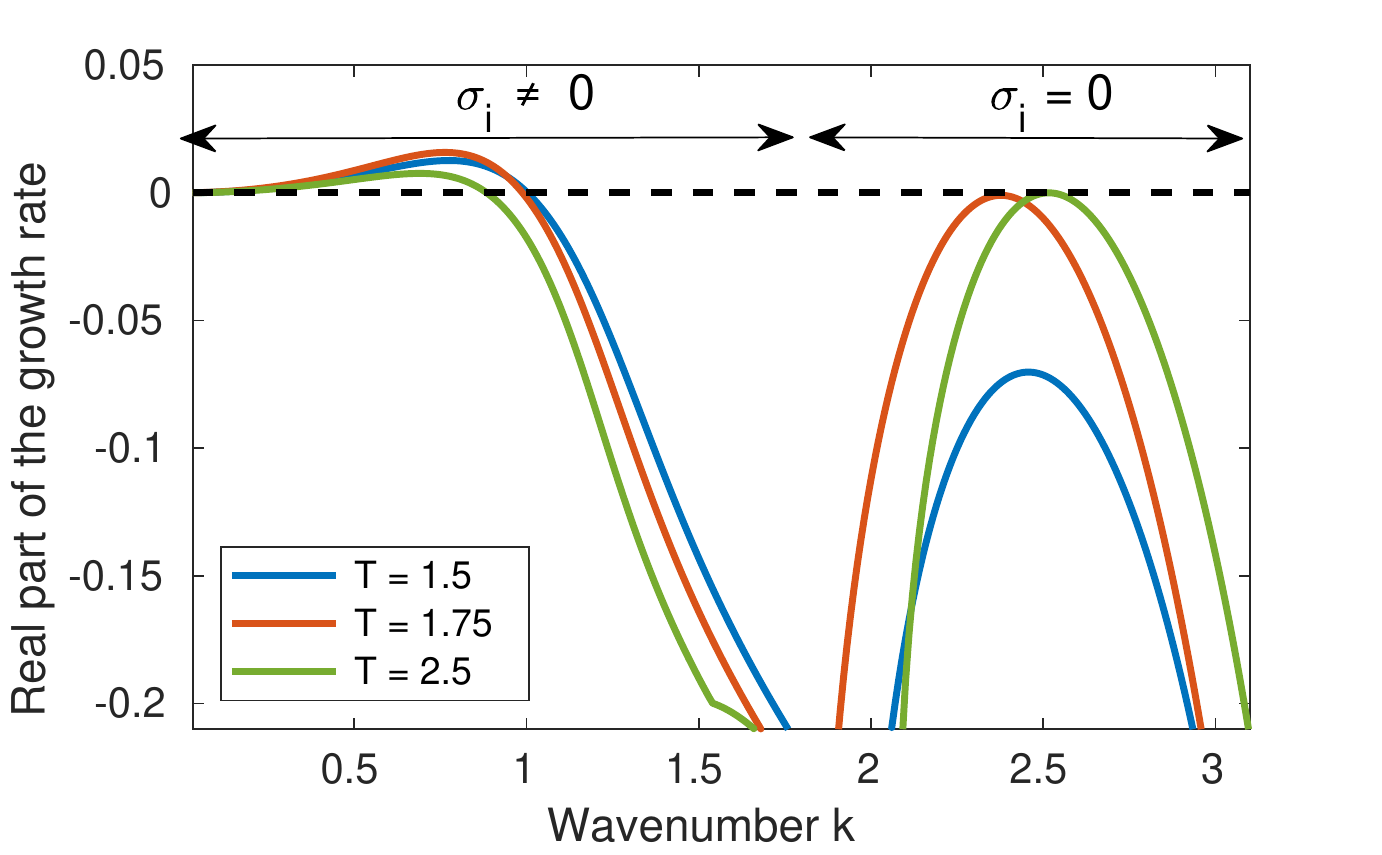}
\end{center}
\caption{DNS, STQL and MTQL simulations of stratified Kolmogorov flow forced from a randomly perturbed rest state with $Re_b = 1$ and $Fr=0.02$. (a) Total energy of the flow as a function of time. For the MTQL simulation, $\sigma_r$ also is plotted. (b) Maximum fluctuation growth rate $\sigma_r$ as a function of $k$ shown for three different times during the MTQL simulation. Note the saturation of $\sigma_r$ to zero at wavenumbers $k\simeq 2.5$ 
and the presence of weakly-amplified {{oscillatory modes (ignored in this study), with non-zero $\sigma_i$, at small $k$}}. 
}\label{Reb1_sigma}
\end{figure}

We first discuss the MTQL dynamics for $Re_b=1$ and $Fr=0.02$.  The simulation is forced from a rest state, i.e. with zero initial velocity and zero buoyancy deviation from the imposed background linear stratification. During the transient phase of the dynamics evident in figure~\ref{Reb1_sigma}a, the real part of the maximum growth rate over all horizontal wavenumbers greater than a minimum viscous threshold (see below) remains negative until $t \simeq 0.175$, when it reaches zero: the amplitude of the fluctuation fields $A$ then jumps to a finite value to prevent further growth of $\sigma_r$. This behavior, which is clearly depicted in the movie included in the {{supplementary material}}, accounts for the discontinuity in the slow-time evolution of the total energy (see insert in figure~\ref{Reb1_sigma}\textit{a}). A nonlinear steady state -- one type of \textit{exact coherent state} (ECS) arising in forced--dissipative nonlinear infinite-dimensional dynamical systems \citep{Kawahara_2001, Lucas_2017, Lucas_2017_2, Parker_2019} -- supported by a \textit{single} horizontal mode with a wavenumber $k = 2.515$ eventually is attained. Note that this single-mode structure is emergent, not imposed.

Figure~\ref{Reb1_sigma}\textit{b} shows the evolution of the growth-rate spectrum during the MTQL simulation. At $t=1.5$, the earliest time depicted, all modes with wavenumbers greater than about unity are damped. {{We have confirmed that the linearly unstable oscillatory modes (with $\sigma_i\ne 0$) arising at small $k$
disappear for vanishingly small stratification; in contrast, stratification is not a necessary condition}} for the instabilities evident at later times for $k\simeq 2.5$, a distinction recently made by \cite{Parker_2019}. {{For simplicity, we choose to avoid constraining the unstable modes arising at small $k$ since $\sigma_r$ remains very small -- and is even smaller at larger values of $Re_b$ (see the inset in figure~\ref{Spectrum_REB}) -- for these modes when the mode with $\mathit{O}(1)$ $k$ is saturated. Nevertheless,}} we emphasize that in principle the MTQL algorithm could be modified to marginalize these modes, too. At an intermediate time, $t=1.75$, a marginal mode with wavenumber $k\simeq 2.25$ corresponds to the mode with locally maximum growth rate, but by $t=2.5$ the wavenumber of this mode has evolved to $k\simeq 2.5$. {{For the given configuration, these modes with $\mathit{O}(1)$~$k$ have zero phase speed (since $\sigma_i=0$) and correspond to Kelvin--Helmholtz instabilities.}}

\begin{figure}
\begin{center}
\includegraphics[width=1\linewidth]{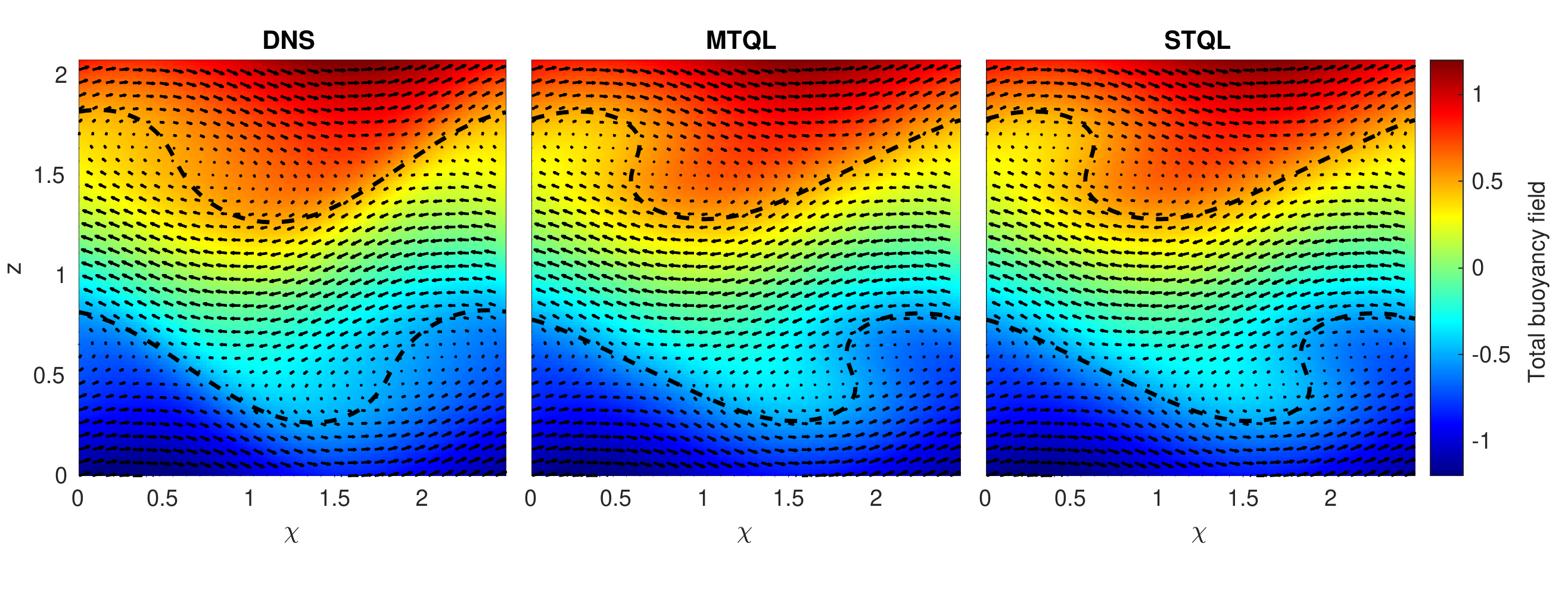}
\end{center}
\caption{Comparison of the steady ECS computed in stratified Kolmogorov flow for $Re_b = 1$ and  $Fr=0.02$ using DNS (left), MTQL (center) and STQL (right) algorithms. The total buoyancy field (including the imposed linear profile) is shown in color, while arrows are used to depict the isotropically scaled velocity field in each case. {{To accentuate differences between the QL results and DNS, the total buoyancy contours $z+\bar{b}=\pm 0.5$ are highlighted (dashed curves).}}}\label{Reb1_plots}
\end{figure}

To perform a quantitative comparison of the three algorithms, both a DNS and STQL simulation also are run for the same parameters ($Re_b=1$ and $Fr=0.02$). For these simulations, the horizontal domain length, $L_x=(2\pi/k)Fr$ and $l_x=2\pi/k$, respectively, is specified using $k=2.515$, i.e. the emergent steady-state wavenumber obtained using the MTQL algorithm. Were it possible to perform the DNS and STQL simulation in the limit $L_x, l_x \rightarrow \infty$, we would expect an ECS with this horizontal wavenumber to be realized. The evolution of the total energy reported in figure~\ref{Reb1_sigma} shows that, in all three simulations, there is an overshoot of the total energy density, corresponding to the ``bursting regime'' documented in \cite{Michel_Chini_2019} and observed in 3D DNS of the full Boussinesq equations  \citep{Rorai_2014,Feraco_2018}. The MTQL and STQL simulations converge precisely to the same steady state, as can be confirmed by inspection of figures~\ref{Reb1_sigma} and \ref{Reb1_plots}; of course, this convergence is not entirely surprising (although not guaranteed) given that the same equations are being solved in the \textit{steady} limit [cf. \eqref{QL_u_bar}--\eqref{QL_bp}].

In figure~\ref{Reb1_plots}, which shows the steady ECS computed using the DNS (left), MTQL (middle) and STQL (right) algorithms, color indicates the total buoyancy field, i.e. including the imposed background linear stratification, while arrows are used to depict the velocity field. To facilitate the comparison of the 2D structure, note that the DNS velocity field $(u, \epsilon^2 W)$ is plotted, while the corresponding fields $(\bar{u}_0+ \epsilon u_1' , \epsilon W_{-1}')$ are shown for the STQL and MTQL simulations [cf. \eqref{UBPexpansions} and \eqref{Wexpansion}]; that is, all velocity components are normalized by $U$. The agreement among the three steady-state ECS lends confidence to the asymptotically-reduced system  \eqref{QL_u_bar}--\eqref{QL_bp} and thereby to the novel MTQL algorithm. The small quantitative discrepancy between the final energies of the ECS obtained from the DNS and via the STQL/MTQL algorithms (figure~\ref{Reb1_sigma}\textit{a}) is partly attributable to the omission of the mean-field correction $\epsilon \overline{u}_1$ from the latter schemes. If desired, this correction could be self-consistently computed by carrying the analysis to higher-order, although we emphasize that the reduced system \eqref{QL_u_bar}--\eqref{QL_bp} is both asymptotically consistent (apart from the diffusive regularization) and closed.

\begin{figure}
\begin{center}
\includegraphics[width=0.7\linewidth]{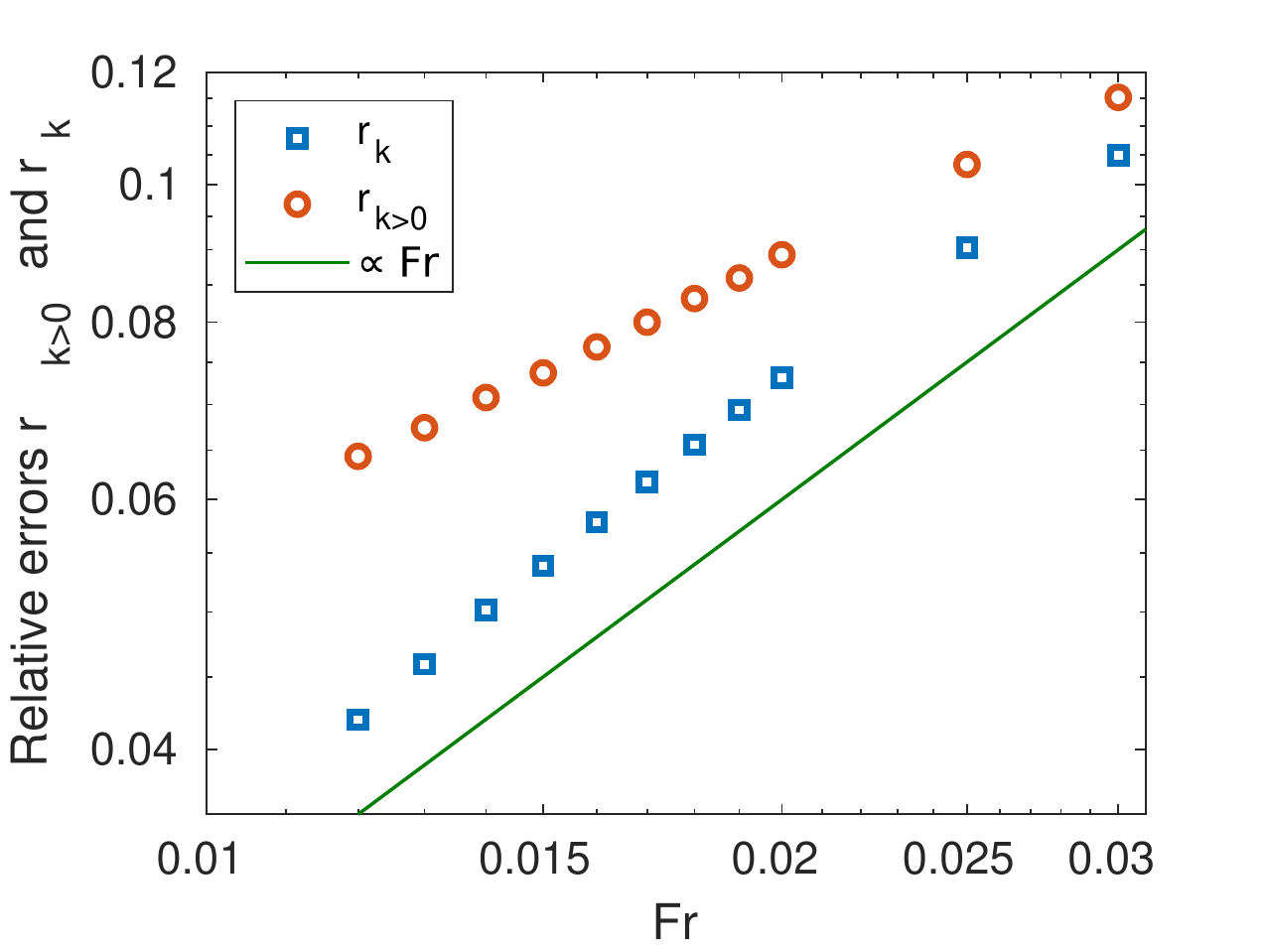}\end{center}
\caption{Convergence of the steady ECS computed at $Re_b$=$1$ using the asymptotically-reduced system \eqref{QL_u_bar}--\eqref{QL_bp} to the corresponding steady solution of the full 2D Boussinesq equations in the limit $Fr \rightarrow 0$. The red symbols show the relative error computed using the total fluctuation energy obtained from the DNS (the sum of the energy in each non-zero horizontal Fourier mode), i.e. $r_{k>0}=|E_{k>0}^{\tiny{\mbox{DNS}}}-E_{k>0}^{\tiny{\mbox{MTQL}}}|/E_{k>0}^{\tiny{\mbox{DNS}}}$. The blue symbols show the relative error evaluated using the fluctuation energy only in the Fourier mode with the fundamental wavenumber $k$, again as obtained from the DNS, $r_{k}=|E_{k}^{\tiny{\mbox{DNS}}}-E_{k}^{\tiny{\mbox{MTQL}}}|/E_{k}^{\tiny{\mbox{DNS}}}$. For reference, the solid green line has a slope equal to $Fr^1$.}
\label{Reb1_convergence}
\end{figure}

The agreement between the DNS and MTQL/STQL simulations is expected to improve as $Fr$ is decreased, since the reduced model is derived in the limit $Fr \rightarrow 0$. To assess the asymptotic convergence of the MTQL algorithm in that limit, a suite of DNS and MTQL simulations with $Re_b=1$ is performed for a set of decreasing values of $Fr$. For the given parameter regime, the MTQL algorithm always converges to a steady state. The long-time dynamics exhibited by the DNS, however, becomes time-dependent for $Fr< 0.012$. Given that the fluctuation-induced mixing is of central importance (see \S~\ref{LargeReb}), we choose to compare, in figure~\ref{Reb1_convergence}, the energy of the fluctuation fields computed using DNS and the MTQL algorithm for $Fr \geqslant 0.012$. (In particular, the contribution of the fluctuations to the total energy is $\mathit{O}(Fr)$, hence the use of this proxy to examine quantitatively the limit $Fr \rightarrow 0$).  Indeed, using Parseval's identity, the dimensionless total energy of the primitive 2D Boussinesq equations (with $\alpha=Fr$), 
\begin{eqnarray*}
E_{\mbox{tot}} = \int \mathrm{d}\chi\mathrm{d}z \left(u^2 + Fr^2 w^2 + b^2\right)/2,
\end{eqnarray*}
can be re-expressed as a sum of the energy contained in the various horizontal modes $k$; that is $E_{\mbox{tot}} = \sum_{k\ge 0}E_k$.

As evident in the figure, the relative error in the fluctuation energy is computed two different ways. The red circles show the relative error when the fluctuation energy is computed using the total DNS fluctuation energy summed across all horizontal Fourier modes with non-zero wavenumber. In addition, the blue squares show the relative error based on the DNS fluctuation energy contained only in the fundamental Fourier mode with wavenumber $k$. Clearly, both metrics confirm that solutions of the asymptotically-reduced equations converge to those of the full Boussinesq equations in the distinguished limit considered, namely, as $Fr\to 0$ with $Re_b$ fixed, although only the data for the fundamental mode approximates the expected rate of decay in the relative error.

\subsection{Exact coherent states for larger $Re_b$}\label{LargeReb}

Steady states can be reached in the MTQL simulations at $Re_b=1$ even for $Fr < 0.012$, a regime in which the DNS 
exhibit persistent unsteady dynamics. Nevertheless, the long-time MTQL dynamics need not be steady or regular on the slow time scale, and for sufficiently large $Re_b$ we anticipate that more complex dynamics would be exhibited  within the two time-scale framework. For the parameter regime considered here, however, the MTQL algorithm -- although ostensibly not specifically designed for this purpose -- has the virtue of yielding \textit{exact coherent states} (ECS) that would be linearly unstable within the full Boussinesq dynamics. For at least the last twenty years, such states have been known to be able to capture certain characteristic attributes of the turbulent regime; see e.g. \cite{Kawahara_2001}. ECS have been of particular interest in the context of wall-bounded constant-density shear flows, where a general mechanism supporting such states has been identified \citep{Hamilton_1995}. Numerical computation of ECS from the governing Navier--Stokes or Boussinesq equations typically requires recurrent flow analyses of expensive direct numerical simulations to provide suitable initial conditions for sophisticated Newton-hookstep solvers, although more efficient approaches have been proposed \citep{Page_2020}. In contrast, it is clear from the present investigation and related studies \citep{Hall_Sherwin_2010,Beaume_2015,Montemuro_2020} that asymptotically-reduced systems, derived using multiple scales analysis, retain the dominant interactions that sustain such states while self-consistently filtering other dynamics, yielding more efficient algorithms for ECS computations and simultaneously exposing the underlying physical mechanisms.

\cite{Lucas_2017} and \cite{Lucas_2017_2} computed ECS in 3D stratified Kolmogorov flow with a \textit{horizontally-varying} forcing (in contrast to the present study). Of particular interest is the mixing efficiency associated with these ECS, which the authors were able to compute for $Re_b \le 95$ and $Fr \ge 0.23$. With a similar aim, we compute by continuation steady-state ECS in 2D stratified Kolmogorov flow using the MTQL algorithm for $Fr=0.01$ and $Re_b\in\left[1,10\right]$; i.e. steady states obtained at given $Re_b$ values are used as initial iterates in MTQL simulations performed at incrementally larger $Re_b$. We again emphasize that the horizontal domain size (or fundamental wavenumber) is not imposed \textit{a priori} but rather is an emergent property of our computations.

\begin{figure}
\begin{center}
\includegraphics[width=0.9\linewidth]{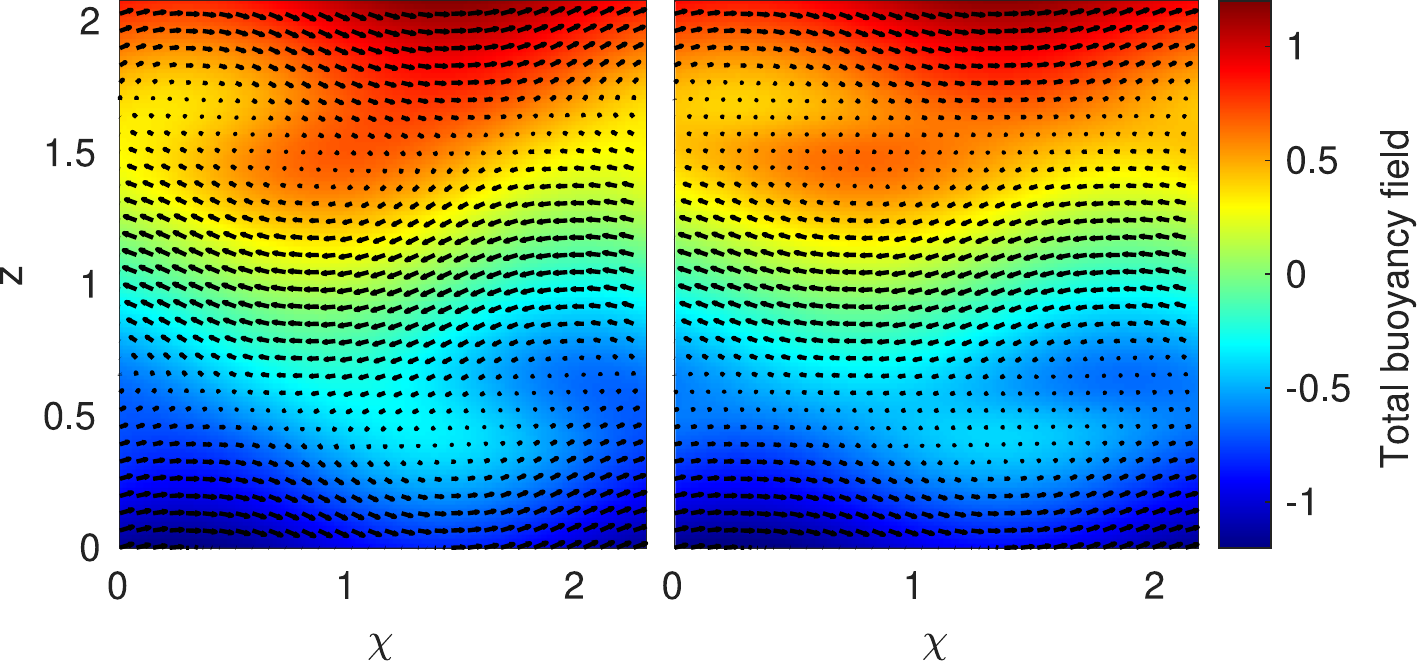}\end{center}
\vspace{0.2cm}
\caption{Comparison of steady ECS in stratified Kolmogorov flow for $Fr=0.01$ and $Re_b = 1$ (left) and $Re_b=10$ (right) computed using the MTQL algorithm. The total buoyancy field (including the imposed linear profile) is shown in color, while the arrows are used to depict the isotropically scaled velocity field in each case.}
\label{Reb_increase_snapshots}
\end{figure}

\begin{figure}
\begin{center}
\includegraphics[width=0.47\linewidth]{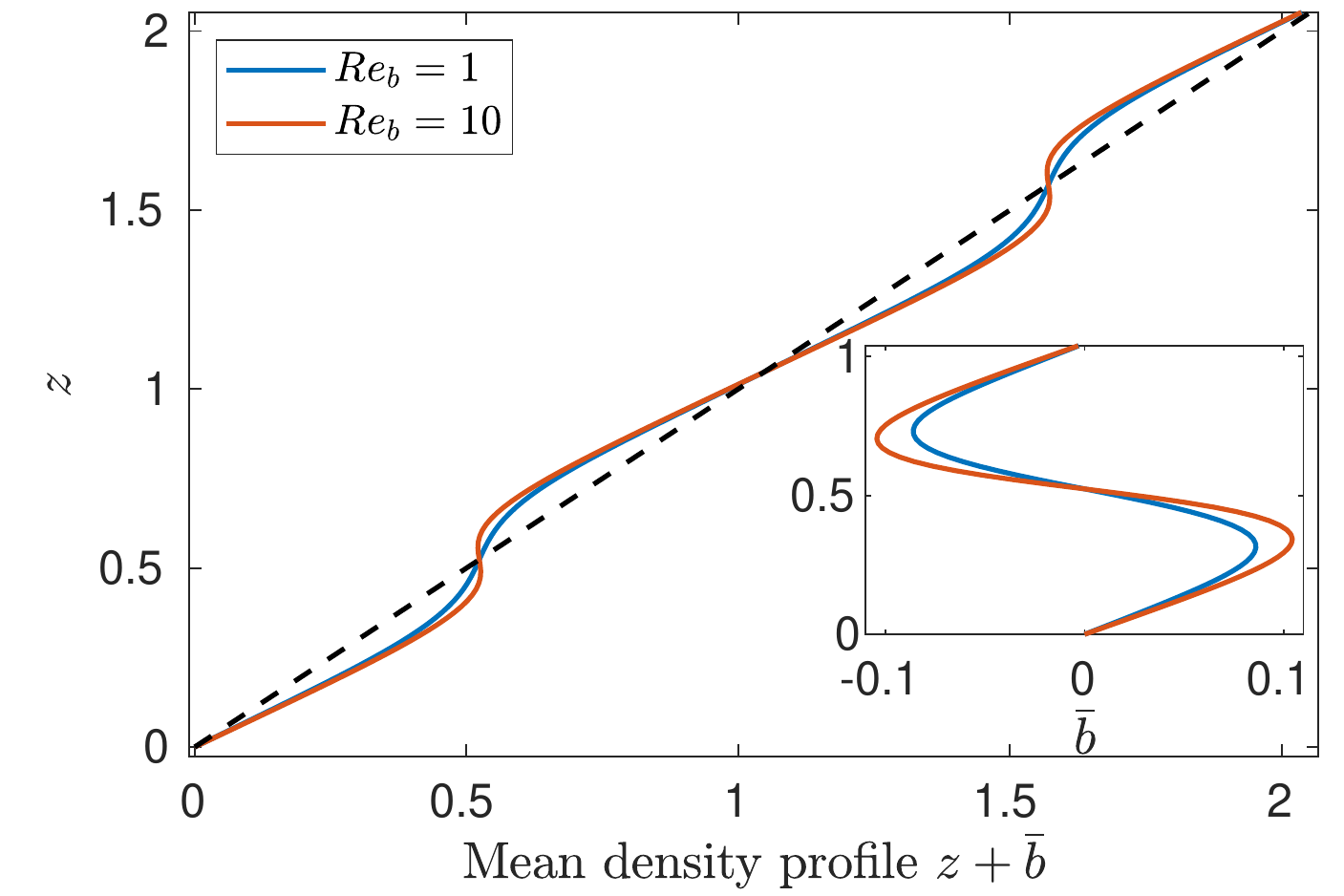}\hspace{0.05\linewidth}\includegraphics[width=0.47\linewidth]{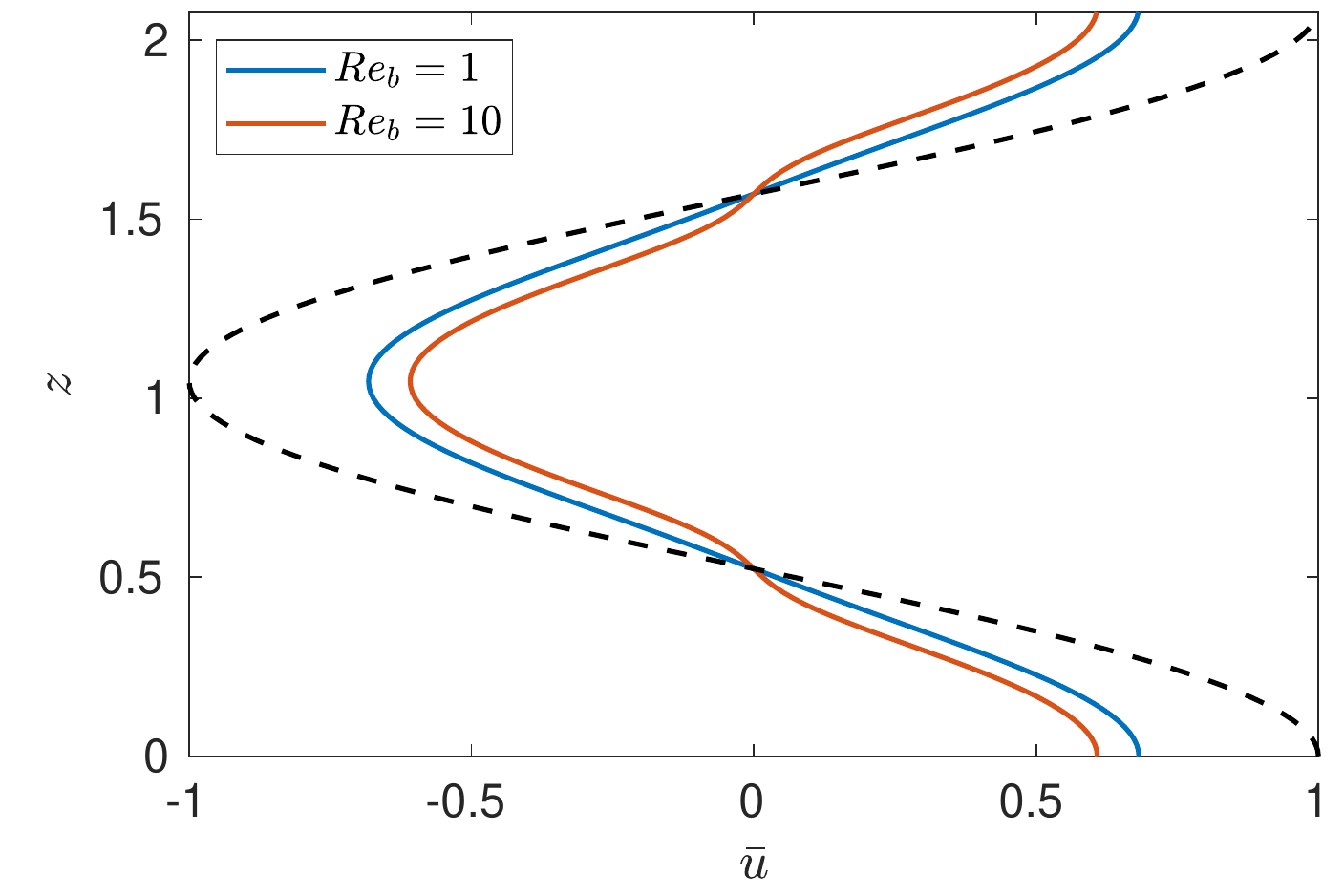}
\end{center}
\caption{Steady-state mean vertical profiles of buoyancy (left) and horizontal velocity (right) computed using the MTQL algorithm for $Fr=0.01$ and $Re_b=1$ (blue) and $Re_b=10$ (red). The dashed lines correspond to the unstable laminar state, with velocity profile $\bar{u}_L=\cos 3z$ and buoyancy perturbation $\bar{b}=0$.}\label{Reb_increase_mean_profiles}
\end{figure}

The steady ECS fields for the two extreme values of $Re_b$ are depicted in figure~\ref{Reb_increase_snapshots}. Even these relatively simple (2D) steady states exhibit features that are commonly associated with stratified turbulence; in particular, the solutions indicate the spontaneous emergence of two layers of relatively well-mixed fluid centered on $z = \pi/6 \simeq 0.5$ and $z = \pi/2 \simeq 1.6$, separated by broader `streams' flowing in opposite directions.
This feature is even more clearly evident in the mean buoyancy and velocity profiles ($\bar{b}$ and $\bar{u}$) shown in figure~\ref{Reb_increase_mean_profiles} and can be interpreted in the light of the linear stability analysis described in \S~\ref{LSA}. The vertical locations of the mixed layers match those of the minima of the gradient Richardson number of the laminar flow; see \eqref{Rigl}. In the inviscid limit, these locations would correspond to critical layers in which $\bar{u}_L-c$ vanishes, leading to discontinuities in both the buoyancy and the vertical derivatives of the horizontal velocity. In practice, the small but finite viscosity smooths these abrupt variations and leads to layers of finite thickness. In the small $Fr$ limit analyzed here, the system \textit{self-adjusts} to an equilibrium state that is approximately marginally stable with respect to the emergent mean profiles -- so the critical layer interpretation remains quantitatively valid provided that $\bar{u}_L$ is replaced with $\bar{u}$, and $z$ (the background stratification) is replaced with $z+\bar{b}$. 

\begin{figure}
\begin{center}
\includegraphics[width=0.475\linewidth]{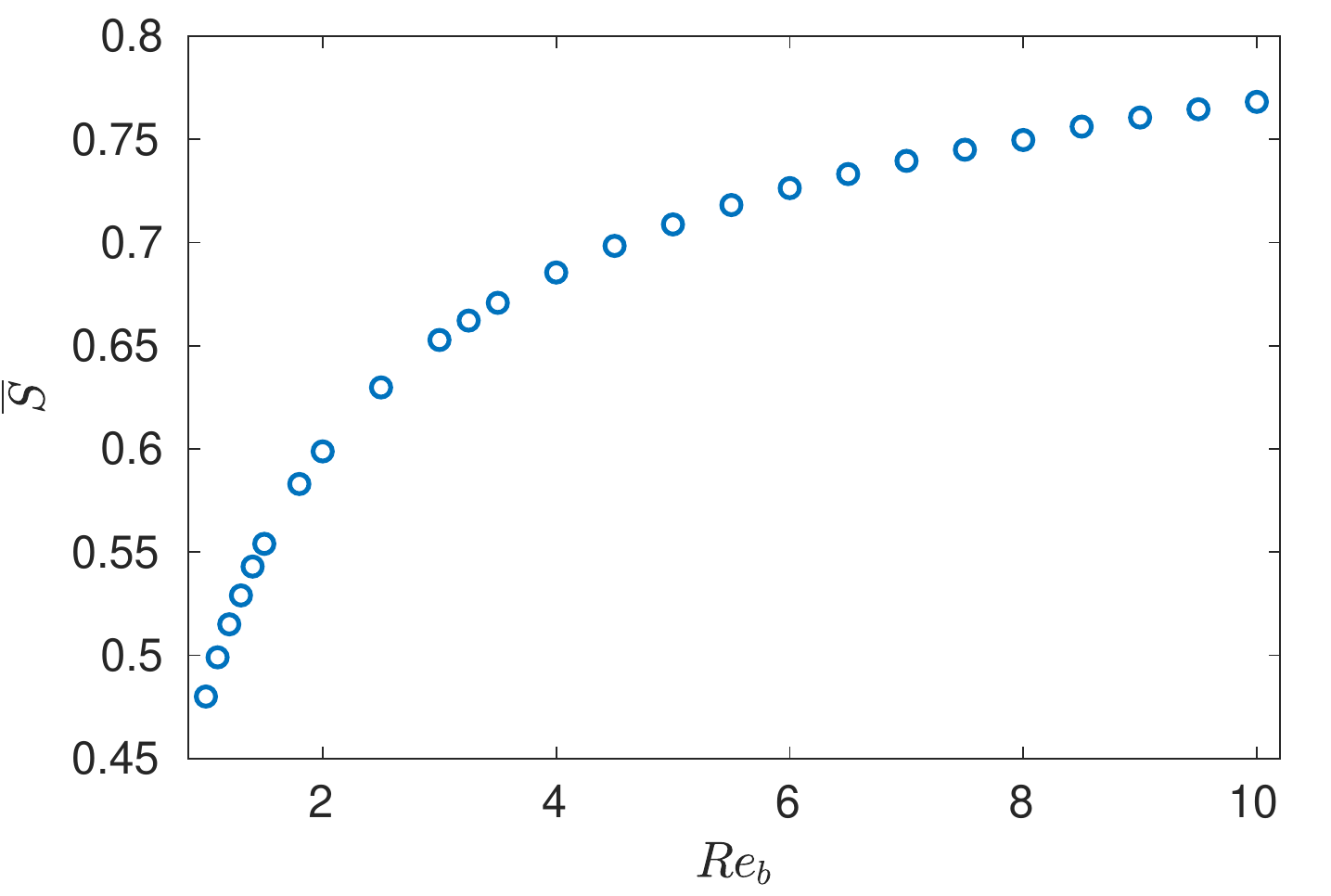}\hspace{0.02\linewidth}
\includegraphics[width=0.485\linewidth]{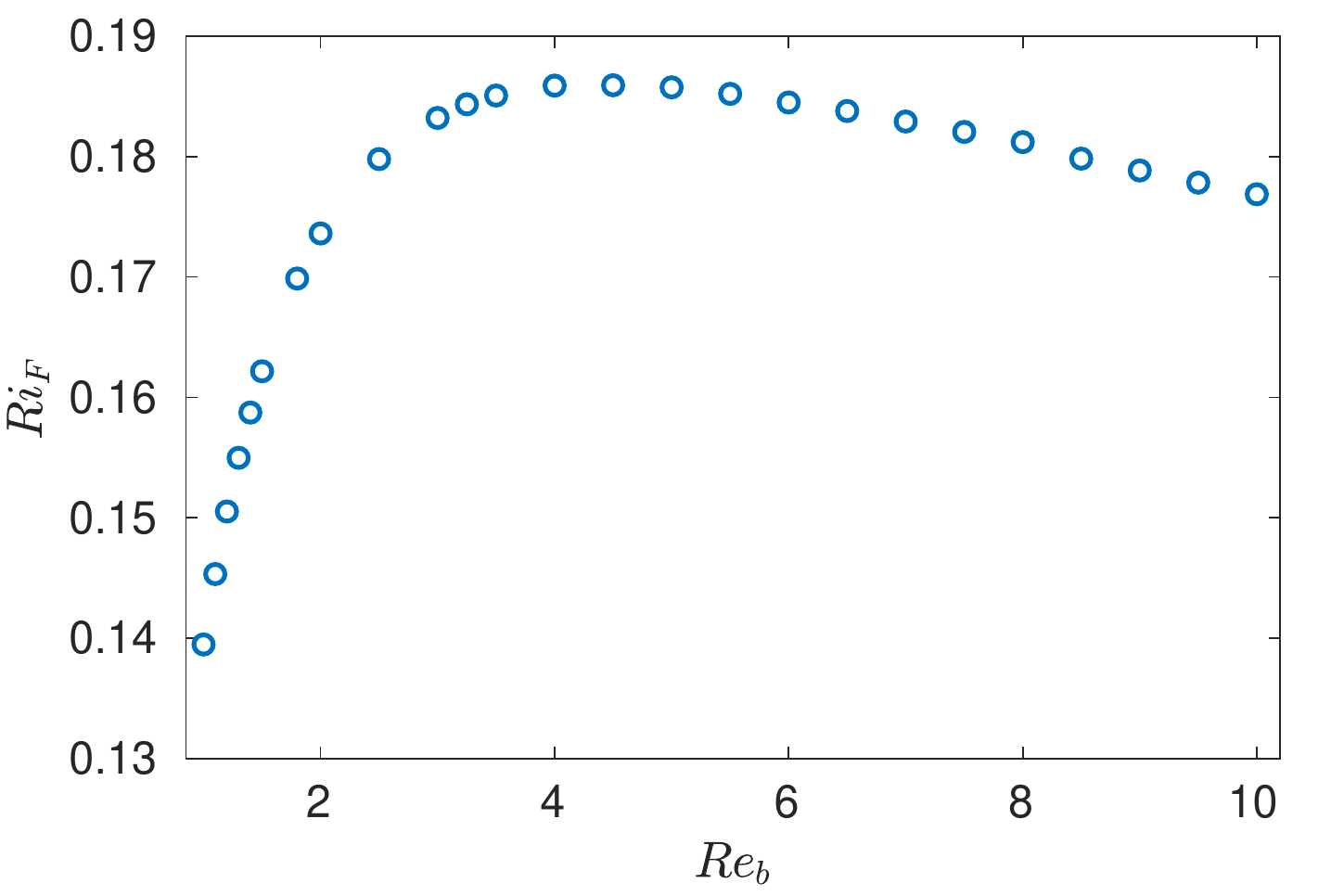}
\end{center}
\caption{Mean quantities computed from the steady state ECS obtained at $Fr=0.01$ as a function of $Re_b$. Left: skewnesses $\bar{S}$. Right: flux Richardson number $Ri_F$.}\label{Reb_mixing}
\end{figure}

The staircase-like profile of $z+b$ has been analyzed thoroughly in the recent 3D DNS study of \cite{Maffioli_2019}, who measured the dimensionless third-order moment (i.e. the skewness)
\begin{equation}
S \equiv  \frac{\left\langle \left(\partial_z b\right)^3 \right\rangle}{\left\langle \left(\partial_z b\right)^2 \right\rangle^{3/2}},
\end{equation}
where $\langle \cdot \rangle$ denotes a spatial average over the entire domain, and it should be recalled that $b(\chi,z) =\bar{b}(z)+b'(\chi,z)$ is the buoyancy deviation from the imposed linear profile ($z$). Interestingly, based on the DNS of \cite{Maffioli_2019}, $S$ does not appear to converge to a finite value in the strongly stratified turbulence regime. Instead, the author finds that $S \sim c Fr^{-0.41}$ (for some constant $c$), the dependence on $Re_b$ not having been investigated. Noting that the layering evident in figure~\ref{Reb_increase_mean_profiles} is related directly to $\bar{b}$ rather than to $b$, we therefore also define the skewness of the horizontally-averaged buoyancy profile
\begin{equation}
\bar{S} =  \frac{\left\langle \left( \partial_z \bar{b}\right)^3 \right\rangle}{\left\langle \left(\partial_z \bar{b}\right)^2 \right\rangle^{3/2}}.
\end{equation}
Indeed, given the expansion \eqref{UBPexpansions}, $S$ reduces to $\bar{S}$ at leading order. The averaging over $\chi$ is, of course, straightforward in the present study owing to the multiple scales decomposition but would prove challenging in a DNS because of the spatial variation of the layer with $\chi$. As shown in figure~\ref{Reb_mixing}, $\bar{S}$ increases monotonically with $Re_b$ with no sign of convergence to a finite value as $Re_b$ is increased. {A crude piecewise linear approximation of the buoyancy profiles yields an estimate of $\bar{S}$ as a function of the ratio of the height of the streams $h_\mathrm{st} \sim L \times Fr $ to that of the emergent mixed layer $h_\mathrm{ML}$ only \citep{Maffioli_2019}. This estimate suggests that $h_\mathrm{ML}/L$ does not become independent of $Re_b$ in the strongly stratified limit; in particular, $h_\mathrm{ML}$ is not simply given by the Ozmidov scale $L_O \sim L \times Fr^{3/2}$.}

Another quantity of interest is the mixing efficiency achieved by these steady ECS. Characterizing stratified mixing in the natural environment is a longstanding question, notably because stratified turbulence in the atmosphere and oceans occurs in a parameter regime in which both $Fr \ll 1$ and $Re_b \gg 1$. As discussed in detail by \cite{Gregg_2018}, an unfortunate additional source of complexity is that several different measures of mixing have been proposed, which can differ significantly from each other in the turbulent regime; see e.g. \cite{Venayag_2016}.  Here, we choose to compute the flux Richardson number $Ri_F$, defined as the  ratio of the volume-integrated buoyancy flux to the power injected by the external force:
\begin{equation}
Ri_F = \frac{\int \tilde{w}\,\tilde{b} ~\mathrm{d}V}{\int \tilde{\mathbf{f}} \boldsymbol{\cdot}  \tilde{\mathbf{u}} ~\mathrm{d}V},\label{Rf}
\end{equation}
where tildes refer to dimensional variables. For steady ECS, this measure coincides with other quantities usually introduced to quantify the mixing, e.g. the mixing efficiency $\mathcal{E}$ based on the dissipation rates of kinetic and available potential energy; see \cite{Caulfield_2000}, \cite{Peltier_2003}, \cite{Salehipour_2015} and \cite{Maffioli_2016_2}. The DNS of \cite{Maffioli_2016_2} and \cite{Portwood_2019} suggest a constant flux coefficient $\Gamma=Ri_F/(1-Ri_F)\simeq 0.2$
in the strongly stratified turbulent limit, whereas the periodic-orbit ECS of \cite{Lucas_2017_2} indicate a progressive decrease toward zero as $Re_b$ is increased. Note that the turbulent flux coefficient
experiences strong variations as a function of $Fr$ in the range $\left[0.05, 0.5\right]$ \citep{Feraco_2018} and that \textit{in situ} measurements in the open ocean indicate a transition from a constant value to a decreasing one at $Re_b \gtrsim 100$ \citep{Lozovatsky_2013,Monismith_2018} although there is still outstanding controversy; see \cite{Caulfield_2020,Caulfield_2021}. DNS in the regime $Fr \ll 1$ and $Re_b \gg 1$ clearly are needed to resolve this issue conclusively but are not feasible with current computing power (figure~\ref{RegimeDiagram}). 

Given the scalings assumed here, the flux Richardson number defined in \eqref{Rf} is computed for the steady-state ECS via
\begin{equation}
Ri_F = - ik |A|^2 \frac{\int_0^{l_z}(\hat{\Psi}^* \hat{b} - \hat{\Psi}\hat{b}^* ) ~\mathrm{d}z}{\int_0^{l_z}\bar{f}\,\bar{u}_0~ \mathrm{d}z}.
\end{equation}
The variation of $Ri_F$ with $Re_b$ is shown in figure~\ref{Reb_mixing} (right). For the range of $Re_b$ considered, $Ri_F$ varies between 0.13 and 0.18. These values correspond to values of the flux coefficient $\Gamma$ 
ranging from approximately 0.16 to 0.22, in reasonable accord with the upper bound on $Ri_F$ of 0.15 proposed by \cite{Osborn_1980} and corresponding to $\Gamma\le 0.2$.
The overshoot of the mixing measure is reminiscent of that reported in both \cite{Maffioli_2016_2} and  \cite{Lucas_2017_2}. Unfortunately, as in other investigations, larger values of $Re_b$ would be required to establish the limiting behavior of $Ri_F$ for these steady ECS. One virtue of the systematically-reduced formulation derived here is that it should facilitate such investigations.  Although we leave this task for a future study, referring to figure~\ref{Reb_mixing} it is conceivable that $Ri_F$ asymptotes to a value near 0.17 for large $Re_b$ (again, for the given steady 2D ECS); intriguingly, this value corresponds to $\Gamma=0.2$, as seen by \cite{Portwood_2019}.

\begin{figure}
\begin{center}
\includegraphics[width=0.75\linewidth]{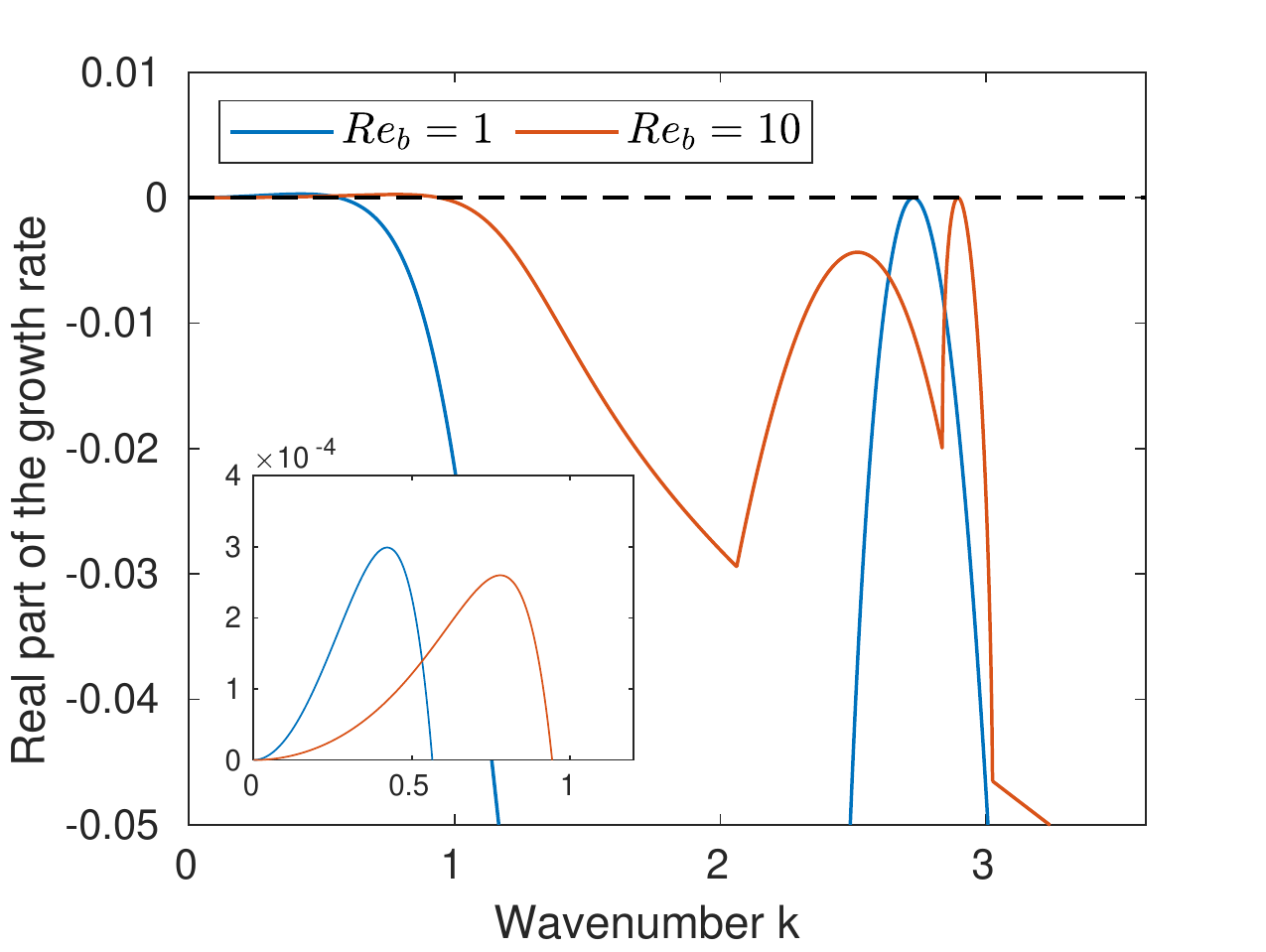}
\end{center}
\caption{Maximum fluctuation-field growth rate as a function of $k$ for the steady ECS computed at $Fr=0.01$ and $Re_b=1$ (blue) and $Re_b=10$ (red). The inset shows the positive growth rates that arise for small $k$; comparing with figure~\ref{Reb1_sigma}$b$ confirms that these low-wavenumber modes are viscous instabilities that vanish as $Re_b\to\infty$.} \label{Spectrum_REB}
\end{figure}

Finally, we discuss the $Re_b$-dependence of the fluctuation growth-rate spectrum  for the steady ECS. As evident in figure~\ref{Spectrum_REB}, the positive linear growth rate regime at small wavenumber is suppressed as $Re_b$ is increased, {{partly justifying our choice to disregard these modes in this investigation since our reduced modeling efforts ultimately are aimed at accessing large values of $Re_b$.}} 
Interestingly, at $Re_b=10$, there is a second local maxima in the growth rate (ignoring the low-wavenumber mode) around $k\simeq 2.5$ that plausibly could spawn the emergence of a second linearly unstable mode for which marginal stability also would have to be enforced. This eventuality could be treated by introducing a second non-zero modal amplitude $B(T)$ and vertical eigenfunctions $[\hat{\Psi}_B(z), \hat{b}_B(z)]$ corresponding to this wavenumber, labeled $k_B$, generalizing the procedure introduced in \S~\ref{MTQL}: instead of the single equation $(\partial \sigma_{Ar}/\partial t)|_{k_A} =\alpha_{Ar} - \beta_{Ar} |A|^2$ for the growth rate $\sigma_{Ar}$ (with associated coefficients $\alpha_{Ar}$, $\beta_{Ar}$) of the sole marginal mode with amplitude $A$, coupled equations for both $\sigma_{Ar}$ and $\sigma_{Br}$ would be derived and solved simultaneously. In this manner, the formalism enables \textit{scale selective adaptivity}, in that modes representing newly emergent length scales are introduced only as needed rather than democratically. Again, this procedure can be implemented with no artificial quantization of wavenumbers imposed by the domain size.

In fact, the potential emergence of a second marginal mode is not the reason we limited this investigation of steady ECS to $Re_b\le 10$. Rather, for $Re_b>10$ with $Fr=0.01$, the steady states become dynamically unstable even within the MTQL algorithm. Ultimately, at a certain time during the evolution, $\sigma_r=0$, $\alpha_r>0$ and $\beta_r<0$.  Physically, this situation corresponds to marginally stable fluctuations that would become linearly unstable (since $\alpha_r>0$) 
and whose feedback on the mean field $\lbrace \bar{u}, \bar{b} \rbrace$ would render the mean field even more unstable (since $\beta_r<0$). In practice, the fluctuation amplitude would reach such large values that the mean field would be forced to respond on the fast time scale, invalidating the \textit{ansatz} of time scale separation (and possibly quasi-linearity). We emphasize, however, that preliminary studies confirm our expectation that this dynamic (a `burst') persists only during a short transient, after which marginal stability is re-established and the MTQL algorithm can be restarted. This issue is described in \cite{Michel_Chini_2019}, and some promising preliminary results involving modification of the MTQL algorithm to account for such bursting regimes are reported in \cite{Ferraro_2019}. Notwithstanding this current limitation, we emphasize that the value $Re_b=10$ is attained with the MTQL algorithm for a realistically small Froude number, \textit{viz.} $Fr = 0.01$. In their 3D DNS, the smallest Froude number reached by \cite{Lucas_2017_2} is $0.008$, for which $Re_b=0.25$, and that by \cite{Maffioli_2016} is $0.02$, for which $Re_b=17$. Although our study has been restricted to 2D stratified Kolmogorov flow, the ECS computations have been performed on a laptop computer, demonstrating the potential for asymptotically reduced modeling of strongly stratified flows with as yet inaccessibly large $Re_b$ and small $Fr$. In contrast, the recent DNS of the 2D Boussinesq equations carried out by \cite{Kumar_2017} required high-performance computing resources to achieve $Re_b=130$ (resp. $Re_b= 300$) for their smallest Froude numbers $Fr=0.16$ (resp. $Fr=0.31$), suggesting that only $O(1)$ values of $Re_b$ could be attained for $Fr=0.01$ with similar computational resources.

\section{Conclusions}\label{CONCLUSION}

Direct numerical simulations consistently show that both local \textit{and} non-local energy transfers occur in strongly stratified turbulence \citep{Waite_2011,Khani_2013,Waite_2014,Augier_2015,Khani_2016,Khani_2018}. Heuristically, these transfers can be understood in terms of the dynamics of the characteristic layered and anisotropic flow structures -- with much larger horizontal than vertical scales -- that are formed owing to the strong stratification. These structures are weakly coupled along the vertical direction, facilitating the occurrence of strong shearing motions. Consequently, in addition to the local energy cascade driven by the self-interaction of these anisotropic structures, stratified shear instabilities can drive non-local energy transfers directly from the large-scale anisotropic flows to much smaller-scale, roughly isotropic motions. While the local energy cascade can be understood using a rescaled variant of Kolmogorov's approach to isotropic 3D turbulence, in particular to predict the energy spectra \citep{Billant_2001,Lindborg_2006}, this approach captures only a subset of the general dynamics. Not only are deviations from these energy spectra reported \citep{Waite_2011,Waite_2014,Augier_2015} but also bursting events, manifested by the non-Gaussianity of the probability density functions of the temperature and vertical velocity \citep{Rorai_2014}, that lead to a mixing enhancement \citep{Feraco_2018}. Both effects may be ascribed to non-local energy transfers.

Guided by these results, we have used multiple scales asymptotic analysis to derive a reduced model of strongly stratified turbulence in the distinguished limit $Fr \rightarrow 0$ and $Re\to\infty$ with $Re_b=Re Fr^2$ fixed, where the buoyancy Reynolds number $Re_b$ emerges as a primary control parameter in the reduced equations that may be systematically increased (numerically) to explore the large $Re_b$ regime. The analysis explicitly recognizes the occurrence of layered anisotropic stratified turbulence (LAST) with dynamics on disparate horizontal and temporal scales. The large scales are characterized by strongly anisotropic velocity layers, with horizontal scales $\mathit{O}(L)$ and vertical scales $\mathit{O}(Fr L) = \mathit{O}(U/N)$ (i.e. the buoyancy scale) and corresponding horizontal velocity $U$ and vertical velocity $Fr U$, that evolve on a slow time scale $L/U$. The analysis confirms that in the small $Fr$ limit, the large-scale dynamics is not only associated with nonlinear interactions with similar flow structures but also with small-scale structures. These latter structures are themselves isotropic, having spatial scales $\mathit{O}(Fr L)$, velocity scales $\mathit{O}(\sqrt{Fr} U)$ and time scale $\mathit{O}(Fr L/U) = \mathit{O}(1/N)$, i.e. comparable with the buoyancy period. A reduced set of equations is then obtained for the fields at both large and small scales. According to the reduced dynamics, the large scales can trigger small-scale (e.g. Kelvin-Helmoltz  or Holmboe wave) instabilities and in return the small scales can exert feedbacks on the large scales through their Reynolds stress and buoyancy-flux divergences. 

A central feature of the reduced dynamics is that the fluctuations satisfy equations that are linear with respect to the mean fields. Herein, this QL reduction is derived self-consistently in a well-defined asymptotic limit rather than being prescribed in an \textit{ad hoc} fashion. Since the mean fields vary slowly in time, the potential exists for the fluctuation fields to grow exponentially while the mean fields remain essentially fixed. Instead, investigation of this and other slow-fast QL systems subject to fast instabilities confirms that, over a wide parameter regime, the system \textit{self-adjusts} to a state in which the mean fields are approximately marginally stable even though the laminar state that would be realized in the absence of instabilities is strongly unstable. Interestingly, this manifestation of self-organized criticality appears to be compatible with recent observations and DNS of stratified shear flows \citep{SmythGRL2013,GeyerJPO2016,SalehipourJFM2018,SmythNature2019} and with the pioneering arguments of \cite{Turner_1973} and \cite{Sherman_1978}. Mathematically, the marginal stability constraint is required to ensure the uniformity of the asymptotic expansions posited in our analysis of strongly stratified shear flows. Following the approach introduced by \cite{Michel_Chini_2019}, the fluctuation dynamics on the fast time scale can be replaced by a linear eigenvalue problem for the vertical structure of the marginal mode(s). The otherwise indeterminate fluctuation amplitude is set by a solvability condition \textit{slaving} the amplitude to the mean fields to ensure that positive growth rates (on the fast time scale) will not be realized once a state of marginal stability is attained. 

Computationally, three primary advantages of this approach accrue. Firstly, the time step for the coupled mean/fluctuation system can be chosen as a fraction of the time scale characterizing the \textit{slow} evolution. By design, this time step is asymptotically much larger than the characteristic time of the fast dynamics. Secondly, the (small) characteristic horizontal length scale of the isotropic fluctuations evolves continuously to ensure that the wavenumber of the fastest-growing mode coincides with that of the marginal mode. Unlike DNS in finite domains, the fluctuation wavenumber is not quantized, effectively capturing the dynamics that would be realized in an arbitrarily large horizontal domain. Finally, our approach naturally enables scale-selective adaptivity, in that additional marginal modes with distinct horizontal wavenumbers can be identified and then introduced only when they are required by the evolving slow dynamics.

To confirm the accuracy and illustrate the merit of the asymptotically-reduced equations and novel multiscale algorithm, a suite of simulations has been performed in the simplified setting of strongly stratified 2D Kolmogorov flow in the absence of slow horizontal variability. Finite-amplitude ECS computed at fixed $Re_b$ are shown to converge to the corresponding steady-state ECS of the full Boussinesq equations as $Fr$ is systematically reduced, confirming the asymptotic validity of the reduced system. In addition, a parametric study of the ECS obtained with the multiscale algorithm for $Fr=0.01$ and increasing values of $Re_b$ ranging from one to ten reveals several interesting features. (Note that DNS in this regime yields persistent time-dependent dynamics.) Firstly, the ECS exhibit spontaneous layering, in which the background linear stratification develops a staircase-like profile with regions of increased stratification (`interfaces') separated by emergent well-mixed `layers'. Although larger values of $Re_b$ would be required to address the limiting behaviors of both the flux Richardson number and the skewness of the horizontally-averaged buoyancy profile, the former varies in a range corresponding to a turbulent flux coefficient $\Gamma$ of approximately 0.2 while the latter suggests that the height of the emergent mixed layers does \textit{not} reduce to the Ozmidov scale in strongly stratified turbulence.

We believe that extension of our multiscale reduced system to incorporate three-dimensional dynamics at higher $Re_b$ ultimately will enable these and other outstanding questions regarding strongly stratified mixing to be addressed. Simultaneously, our approach opens new avenues for more accurate sub-grid-scale parameterizations, e.g. to be used in hydrostatic general circulation models, than currently can be achieved with variants of eddy-viscosity modeling. Several mathematical and algorithmic advances, however, are needed. Most importantly, the transient \textit{fast} dynamics of the large scales must be properly captured: during these short periods, temporal scale separation is lost at least intermittently. {{Reincorporation of slow horizontal spatial variability similarly is necessary, but raises attendant questions about the numerical treatment of the local downscale energy cascade that would be generated within the context of the multiscale system.}} A mathematical formalism for predicting the evolution of the wavenumber of the marginal mode not only would be more elegant but presumably more efficient than the rather brute-force approach implemented here. Finally, the reduced model itself could be improved by incorporating a second \textit{vertical} scale to more properly account for the potentially complex dynamics that may be realized within the thin, emergent mixed layers. If feasible, this advance would obviate the need for reintroducing the Froude number $Fr$ into the multiscale reduced system and may capture new dynamical phenomena associated with nonlinear critical layer dynamics. This extension also would be compatible with the emerging conceptual picture that the turbulent motions in apparently very strongly stratified turbulence, with small (global) values of $Fr$, occupy only a small fraction of the total stratified fluid volume -- and in those regions, the stratification is locally very much eroded, as identified using a robust automatic algorithm by \cite{Portwood_2016}. All these challenges are the subject of ongoing research \citep[e.g. see][]{Ferraro_2019}.\\

\noindent
\textbf{Funding.} GPC and GM acknowledge the hospitality of the Kavli Institute for Theoretical Physics, where much of this research was completed and  supported in part by the National Science Foundation under Grant No. NSF PHY-1748958. GPC also gratefully acknowledges support from the Office of Naval Research under Grant No. ONR-BRC N000141712307. All authors acknowledge support from the Woods Hole Summer Program in Geophysical Fluid Dynamics, where this work was initiated. \\
\noindent
\textbf{Declaration of Interests.} The authors report no conflict of interest.
\bibliographystyle{arfmtest}

\end{document}